# Circuit Transformations for Quantum Architectures

Andrew M. Childs[*,1,2,3], Eddie Schoute[†,1,2,3], and Cem M. Unsal[4]

[1]Joint Center for Quantum Information and Computer Science, University of Maryland
[2]Institute for Advanced Computer Studies, University of Maryland
[3]Department of Computer Science, University of Maryland
[4]Department of Mathematics, University of Maryland

September 6, 2019

**Abstract**

Quantum computer architectures impose restrictions on qubit interactions. We propose efficient circuit transformations that modify a given quantum circuit to fit an architecture, allowing for any initial and final mapping of circuit qubits to architecture qubits. To achieve this, we first consider the qubit movement subproblem and use the Routing via Matchings framework to prove tighter bounds on parallel routing. In practice, we only need to perform partial permutations, so we generalize Routing via Matchings to that setting. We give new routing procedures for common architecture graphs and for the generalized hierarchical product of graphs, which produces subgraphs of the Cartesian product. Secondly, for serial routing, we consider the Token Swapping framework and extend a 4-approximation algorithm for general graphs to support partial permutations. We apply these routing procedures to give several circuit transformations, using various heuristic qubit placement subroutines. We implement these transformations in software and compare their performance for large quantum circuits on grid and modular architectures, identifying strategies that work well in practice.

## 1 Introduction

Quantum algorithms are typically formulated in a circuit model in which two-qubit gates can be performed between any pair of qubits. However, most realistic quantum architectures impose restrictions on qubit interactions. Thus a natural challenge is to find a way of implementing a given circuit on a given architecture with low overhead. We can do this by finding a time-efficient *architecture-respecting circuit transformation*—a mapping to a new circuit that preserves the function of the original quantum circuit up to an initial mapping of circuit qubits to architecture qubits and a final mapping of architecture qubits back to circuit qubits, where the new circuit is constrained to respect the architecture.

There have been many proposals for the design of quantum processors. Examples include trapped ion systems that enable interactions between any two ions in a trap [40] and superconducting qubit architectures with more limited interactions [20, 50, 45]. Many proposed architectures for scalable devices employ modularity, building a large device from interconnected subunits. For example, one proposal considers registers of ion trap qubits coupled via photonic quantum channels through a reconfigurable optical switch [40, 41]. Another approach uses shielded modules of circuit quantum electrodynamics devices connected by superconducting transmission wires [12].

---

[*]amchilds@umd.edu
[†]eschoute@cs.umd.edu

There is also a considerable amount of work on implementing circuits under architectural constraints. Some examples include implementations of Shor's algorithm [18], the quantum Fourier transform on 1D nearest-neighbor architectures [36], and quantum adders on nearest-neighbor architectures [16, 15]. However, the aforementioned works focus on analyzing specific circuits. Instead, we wish to find automated circuit transformations that can handle complex circuits and compare their performance when implemented in various architectures. Bounds on the efficacy of architecture-respecting circuit transformations and good automated tools for implementing them may be able to inform architecture design decisions [54]. Unfortunately, it is challenging to achieve good performance with an automated tool. Indeed, finding even one optimal placement for a set of gates is NP-hard [35].

**Prior Work on Automated Architecture-Respecting Circuit Transformations** Several previous works use exhaustive approaches that take time exponential in the number of qubits (and hence can only be used for small instances). For example, Saeedi, Wille, and Drechsler [47] use SAT solvers to decompose circuits so they can be run on the path architecture; [34] finds an optimal circuit transformation on nearest-neighbor architectures by formulating the problem as a pseudo-boolean optimization; Venturelli et al. [52] use temporal planners to schedule gates; and [42] uses satisfiability modulo theory solvers to find mappings of the circuit with high success probability using calibration data. Other work has instead proposed minimizing the distance between all qubits in groups of gates on specific architectures [49, 56, 44], but this is also NP-hard in general. These and other papers add SWAP gates so that the logical state of a given physical qubit is transferred to a different physical qubit (henceforth, we simply refer to this as *qubit movement*, with the implicit understanding that only the logical state is moved).

As a heuristic solution, we can break the circuit into sets of disjoint gates and move the qubits between each set. Metodi et al. [37] propose polynomial-time heuristic routines that prioritize gates with many dependents. Hirata et al. [23] propose exhaustive and heuristic searches for good placements of qubits on the path architecture and use those to construct circuit transformations.

One can also use heuristic qubit placement and movement algorithms on fault-tolerant 2D grid architectures [31] or algorithms that are designed to handle the surface code [29]. We do not consider fault tolerance explicitly and instead work only at the logical level.

An exhaustive search of all permutations of $n$ qubit locations takes time $O(n!)$ but can work well for small numbers of locations [55], or can be done selectively using $A^*$ heuristic search [59, 60] or local search [35, 8]. By choosing a suitable initial placement of qubits, we can further reduce the qubit movement cost. For example, [30] tries to find a good initial placement by repeatedly transforming the quantum circuit forwards and then backwards, taking the output qubit placement as input for the next iteration.

Other work has considered a model in which one can perform fast measurements and adapt later parts of the computation based on the outcomes [21]. This model allows the movement of qubits with just a constant overhead at the cost of extra ancillas [46]. However, realizing such a model presents significant technical challenges and we do not consider it here.

Various bounds are also known for the cost of moving qubits. Sorting networks provide a way to upper bound the depth of the qubit movement circuit [28, 7, 13]. Further, there exists a circuit where the depth overhead of qubit movement is at least logarithmic for architectures with finite degree [22]. We refer to [3] for a more complete overview of sorting networks.

**Contribution** In this paper, we construct architecture-respecting circuit transformations that attempt to minimize the circuit depth or size overhead and have worst-case time complexity polynomial in the sizes of the circuit and architecture graph. We model the connectivity of the underlying hardware as a simple graph where vertices represent the qubits and edges represent places where a two-qubit gate can be performed.

As a simple and fast approach, we propose the *greedy swap circuit transformation* (Section 2.2.2). It inserts SWAPs on edges chosen to minimize the total distance between qubits involved in two-qubit gates until some gate(s) can be executed.

We then propose building architecture-respecting circuit transformations (Section 2.2.3) by combining algorithms for two basic subproblems: qubit movement (addressed by *permuters*, for which we provide theoretical performance guarantees) and qubit placement (addressed by *mappers*). For the latter, we specify a variety of heuristic strategies (Section 4) to find suitable placements of qubits from the input circuit, attempting to optimize for circuit size or depth. We implement these algorithms in software, which is publicly available under a free software license [48].

Consider now the problem of moving qubits on a given architecture graph. A sorting network sorts any fixed-length sequence of integers with a circuit of comparators, which compare two inputs and output them in nondecreasing ordering. While sorting networks can be used to route qubits [7], they achieve a more general task, and the cost of routing can sometimes be lower with other methods. Specifically, we suggest ROUTING VIA MATCHINGS [2] (introduced in Section 3.1) as a more suitable framework for moving qubits in parallel. Deciding whether there exists a depth-$k$ circuit for ROUTING VIA MATCHINGS is NP-complete in general for $k > 2$ [3], but optimal or near-optimal protocols are known for specific graph families [2, 58]. In some cases it is possible to implement any permutation asymptotically more efficiently than a general sorting network (see Table 1). On complete graphs, for example, any permutation can be implemented in a depth-2 circuit of transpositions [2], whereas an optimal sorting network has depth $\Theta(\log n)$ [1].

While it is common to consider only the worst-case routing performance, we also wish to route efficiently in practice. To improve practical performance, we generalize to *partial permutations* (permutations only defined on some subdomain) so that we can also move subsets of qubits efficiently. The destinations of the remaining qubits are unconstrained. In Section 3.1, we present routing algorithms for the path graph, the complete graph, and the *generalized hierarchical product* of graphs [6], which includes the Cartesian product of graphs and *modular* architectures as special cases [41]. Graphs obtained as hierarchical products have many good properties for quantum architectures [5]. We establish an upper bound on the routing number of a hierarchical product (Theorem 3.4) that matches prior work for total permutations on the Cartesian product of graphs [2] and depends on easily computable properties of the input partial permutation.

We also propose using TOKEN SWAPPING [57] for minimizing the total number of SWAPs, which is relevant when optimizing for total circuit size (Section 3.2). We generalize this problem to partial permutations and obtain a 4-approximation algorithm (Theorem 3.10).

Finally, we evaluate our circuit transformations on large quantum circuits (Section 5) and compare their performance with the circuit transformation included in the Qiskit software (Section 2.2.1) [8]. We find that the relative performance varies significantly with the circuit type and architecture. When minimizing circuit size, the greedy swap circuit transformation is one of the best, though some improvement may be gained using some of our specialized circuit transformations. For depth, some of our specialized circuit transformations do best on random circuits on grid architectures, whereas Qiskit's circuit transformation does well on modular architectures. For quantum signal processing circuits [33] we find that the depth is best minimized by our greedy swap circuit transformation.

|  | Worst-case circuit depth | |
| --- | ---: | ---: |
| Graph family | Sorting (comparators) | Routing nr. (transpositions) |
| path ($P_n$) | $n$ [25] | $n$ [2] |
| complete ($K_n$) | $\Theta(\log n)$ [1] | 2 [2] |
| tree with max degree $\Delta$, diameter $D$ | $O(\min(\Delta, \log(n/D))\, n)$ [4] | $3n/2 + O(\log n)$ [58] |
| gen. hierarchical product, $\Pi_{\mathbf{v}}(G_1, G_2)$ | not known | $\left\lceil \frac{|V_2|}{\text{ham}(\mathbf{v})} \right\rceil (\text{rt}(G_1) + \text{rt}(G_2)) + \text{rt}(G_2)$ |
|   Cart. product $G_1 \times G_2 = \Pi_{\mathbf{1}}(G_1, G_2)$ | not known | $2\,\text{rt}(G_1) + \text{rt}(G_2)$ [2] |
|   $r$-dimensional grid $\left(\bigtimes_{i=1}^{r} P_{n_i}\right)$ | $n_1 + 2\sum_{i=2}^{r} n_i + o(\cdot)$ [27] | $n_1 + 2\sum_{i=2}^{r} n_i$ [2] |
|   modular architecture, $\Pi_{\mathbf{e}_1}(K_{n_1}, K_{n_2})$ | not known | $3n_2 + 2$ |

**Table 1:** Performance bounds for sorting networks versus routing via matchings (the routing number, $\text{rt}(G)$; see (9)) where $|V| = n$. Let $G_2 = (V_2, E_2)$ and $\mathbf{v} \in \{0, 1\}^{|V_2|}$; $\text{ham}(\mathbf{v})$ is the Hamming weight of $\mathbf{v}$, $\mathbf{1} \coloneqq [1 \ldots 1]$, and $\mathbf{e}_1 \coloneqq [1\, 0 \ldots 0]$. We list special cases of the generalized hierarchical product (see Definition 3.2): the Cartesian product of graphs, the $r$-dimensional grid, and the modular graph. See [4] for a short overview of known lower and upper bounds for sorting networks restricted to common topologies.

## 2 Constructing Circuit Transformations

Program transformations are algorithms that modify computer programs while retaining functionality [43]. In a similar vein, we define a *circuit transformation* as an algorithm that modifies an input quantum circuit to produce an output quantum circuit with the same functionality. We represent an architecture by a simple graph $G = (V, E)$, and let $Q$ denote the set of qubits of the input circuit. A circuit transformation is *architecture-respecting* if it produces injective initial and final mappings of the form $\hat{p} \colon Q \to V$ and an architecture-respecting output circuit. The output circuit is architecture-respecting if for each two-qubit gate acting on (qubit) vertices $v_1, v_2$ we have $(v_1, v_2) \in E$ (where the ordering is irrelevant since $G$ is undirected). Henceforth, we only consider circuit transformations that are architecture-respecting, and we refer to them simply as circuit transformations. We propose a construction for a general circuit transformation that may use the properties of the underlying architecture by relying on a specialized subroutine for moving qubits called a *permuter* (Section 3), and a subroutine determining where to place qubits, called a *mapper* (Section 4).

To be able to transform a circuit, we must have $|Q| \leq |V|$, and the output circuit must contain a qubit for every vertex in the architecture. Throughout the circuit transformation, we keep track of the injective current placement of qubits $\hat{p} \colon Q \to V$. The initial and final values of $\hat{p}$ are also the initial and final mappings, respectively, of qubits to the architecture. A gate is *executed* by appending it to the output circuit. Two-qubit gates with qubits $q_1, q_2 \in Q$ can only be executed when $(\hat{p}(q_1), \hat{p}(q_2)) \in E$. By adding SWAP gates to the output circuit, we can change $\hat{p}$ and thereby unitarily transform quantum circuits for execution on an architecture.

### 2.1 Definitions

We define some terminology that will be used throughout the paper.

#### 2.1.1 Partial Functions and Partial Permutations

For sets $X$ and $Y$, a *partial function* $f \colon X \rightharpoonup Y$ is a mapping from $\text{dom}(f) \subseteq X$ to $\text{image}(f) \coloneqq \{f(x) \mid x \in \text{dom}(f)\} \subseteq Y$. However, $f(x)$ is undefined for $x \in X \setminus \text{dom}(f)$. We consider such elements $x$ *unmapped*. For $x \in \text{dom}(f)$, we write $x \mapsto f(x)$ and say that $x$ is *mapped to* $f(x)$. We can then define any partial function $f$ as a set of mappings, $f := \{x \mapsto y \mid x \in X, y \in Y\}$, where all preimages must be distinct (i.e., if $x \mapsto y \in f$ and $x' \mapsto y' \in f$ with $y \neq y'$, then $x \neq x'$). A *total* function $\hat{f}$ is a partial function where $\text{dom}(\hat{f}) = X$ and is denoted $\hat{f} : X \to Y$. By the term "function" we will mean a total function.

A partial function $f$ is *injective* iff $\forall x, x' \in \text{dom}(f)$ with $x \neq x'$, $f(x) \neq f(x')$. A function $\hat{f} : X \to Y$ is *surjective* iff $\forall y \in Y, \exists x \in X : f(x) = y$. A *bijective partial function* $f$ is a partial function that is injective and is denoted $f : X \rightharpoonup Y$ (note that such an $f$ is necessarily surjective on its image). A *bijective function* $\hat{f}$ is both injective and surjective and is denoted by $\hat{f} : X \leftrightarrow Y$. For any bijective (partial) function $f$ there exists an inverse function $f^{-1} : \text{image}(f) \to \text{dom}(f)$.

A partial permutation $\pi$ is any bijective partial function with the same domain and codomain, i.e., $\pi : X \rightharpoonup X$. Similarly, a total permutation is any $\sigma : X \leftrightarrow X$. By "permutation" we mean a total permutation.

We also define some notions specifically useful for this paper. An *unmapped vertex* is a vertex in $V \setminus \text{dom}(\pi)$, for a graph $G = (V, E)$ and $\pi : V \rightharpoonup V$. We define the union of partial functions $f : X \rightharpoonup Y$ and $g : X \rightharpoonup Y$ when $\text{dom}(f) \cap \text{dom}(g) = \emptyset$ as

$$(f \cup g)(x) := \begin{cases} f(x) & \text{if } x \in \text{dom}(f), \\ g(x) & \text{if } x \in \text{dom}(g). \end{cases} \tag{1}$$

Furthermore, $(f \cup g)$ is a bijective partial function iff $f$ and $g$ are bijective partial functions and $\text{image}(f) \cap \text{image}(g) = \emptyset$. A *completion* of $\pi : X \rightharpoonup X$ is a $\hat{\pi} : X \leftrightarrow X = (\pi \cup \sigma)$ for some $\sigma : X \rightharpoonup X$, where $\text{dom}(\sigma) = X \setminus \text{dom}(\pi)$ and $\text{image}(\sigma) = X \setminus \text{image}(\pi)$.

### 2.1.2 Directed Acyclic Graph Representation of a Circuit

A quantum circuit can be viewed as a directed acyclic graph (DAG), where vertices represent gates and directed edges represent qubit dependencies. We define the first *layer* of the DAG, $L$, to be the set of all vertices without predecessors. By removing $L$ and taking the first layer of the resulting DAG, we can define the second layer, and so on.

The size of a circuit is the number of gates it contains (i.e., the number of vertices in the DAG); the depth of a circuit is the number of layers. It is natural to minimize either the depth (Section 4.1), corresponding to the execution time when gates can be applied in parallel, or the the size (Section 4.2), corresponding to the total number of operations that must be performed. We are mainly interested in two-qubit gates and their qubits. Thus we define $\text{tg} : V_D \to Q \times Q$, where $V_D$ is the set of DAG vertices corresponding to two-qubit gates, that outputs the pair of qubits acted on by a given gate. For simplicity, we denote $\text{tg}(L) := \{\text{tg}(g) \mid g \in L, g \text{ is a two-qubit gate}\}$.

## 2.2 Architecture-Respecting Circuit Transformations

We now describe some specific architecture-respecting circuit transformations. We first describe two basic circuit transformations, one provided by the Qiskit software (Section 2.2.1) and another that uses a simple greedy approach (Section 2.2.2). Then, in Section 2.2.3 we specify a family of circuit transformations that builds on specialized procedures for qubit placement and routing.

### 2.2.1 Qiskit Circuit Transformation

The open-source quantum computing software framework Qiskit [8] contains a circuit transformation[1] that we build upon in one of our approaches (Section 4.2.4). We specify this transformation here and compare it with our other approaches to circuit transformations in Section 5.

We initialize $\hat{p}$ arbitrarily. Fix a number of trials, $k \in \mathbb{N}$, for each layer. We do the following in trial $i \in [k]$ where $[k] := \{1, \ldots, k\}$: For all $v, u \in V$, sample a symmetric weight

$$d_i(v, u) = (1 + \mathcal{N}(0, 1/N)) \, d(v, u)^2 \tag{2}$$

independently for $(v, u) \in V \times V$, where $\mathcal{N}(\mu, \sigma)$ represents a sample from the normal distribution with mean $\mu \in \mathbb{R}$ and standard deviation $\sigma \geq 0$, and $d : V \times V \to \mathbb{N}$ is the shortest distance function on the architecture graph. We define an objective function as the sum of gate distances,

$$S := \sum_{(q_1, q_2) \in \text{tg}(L)} d_i(\hat{p}(q_1), \hat{p}(q_2)). \tag{3}$$

We now try to SWAP pairs of qubits to decrease $S$. Specifically, we construct a set of SWAPs by iterating over all edges $e \in E$ and greedily adding the corresponding SWAP if it decreases $S$ and neither endpoint of $e$ is already involved in some SWAP. We execute the set of SWAPs and update $S$. We then iterate this process until either $S = |\text{tg}(L)|$; or there is no SWAP that decreases $S$; or we reach the upper bound of $2|V|$ iterations.

Now, if $S = |\text{tg}(L)|$ then the algorithm has successfully found a sequence of SWAPs and all gates in $L$ can be executed. The result of trial $i$ is then set to this sequence of SWAPs. Otherwise, trial $i$ is a failure. If there is at least one successful trial out of $k$ trials, we execute the SWAPs of a successful trial with the fewest SWAPs and then execute all gates in $L$.

If no trial was successful, we apply the same routine for finding SWAPs that minimize $S$, but taking only a single gate $(q_1, q_2) \in \text{tg}(L)$ at a time. Note that this results in a sequence of SWAPs along the shortest path between $\hat{p}(q_1)$ and $\hat{p}(q_2)$. After each such step we execute the selected gate. We repeat this until all gates in $\text{tg}(L)$ have been executed and also execute all single-qubit gates in $L$. Finally, we remove the vertices in $L$ from the input circuit DAG and iterate this process until all gates in the input circuit are executed.

### 2.2.2 Greedy Swap Circuit Transformation

We also describe a simple greedy approach to circuit transformations. Similar to the Qiskit circuit transformation described above, we prioritize SWAPs that maximally reduce the total distance between the qubits $\text{tg}(L)$, but now using the simpler objective function

$$R := \sum_{(q_1, q_2) \in \text{tg}(L)} d(\hat{p}(q_1), \hat{p}(q_2)). \tag{4}$$

Note that this is different from (3), where a randomized distance $d_i$ is used.

We construct an initial $\hat{p}$ as follows. Let us consider the first layer $L'$ of the circuit consisting of only two-qubit gates (i.e., single-qubit gates are ignored), initialize $p' : Q \rightharpoonup V$ as undefined everywhere, and set $U := \emptyset \subseteq V$. We iteratively construct

$$p' \leftarrow p' + \{q_1 \mapsto v_1, q_2 \mapsto v_2 \mid (q_1, q_2) \in L', (v_1, v_2) \in M\}, \tag{5}$$

---

[1] We base our description on `qiskit.mapper.swap_mapper` from Qiskit version 0.6.1.

where $M \subseteq E$ is a maximum matching of $G$, remove $(q_1, q_2)$ from $L'$, set $U \leftarrow U \cup \{v_1, v_2\}$, and recompute $M$ on the subgraph of $G$ with the vertices $V \setminus U$.[2] The remaining qubits $Q \setminus \text{dom}(p')$ are arbitrarily mapped to the available vertices $V \setminus \text{image}(p')$ to obtain $\hat{p}$.

In every iteration, we construct a set of disjoint gates to execute. We first execute as many gates from $L$ as possible given $\hat{p}$, and we remove these gates from the input circuit. Second, let $E_i$, for $i \in [2]$, be the set of edges where executing a SWAP would decrease $R$ by $i$, excluding edges which already had a vertex involved in a gate this iteration. We then greedily execute gates from $E_2$ first and $E_1$ second, updating both $E_i$s as we go. If we were not able to execute a gate from $L$ and no SWAPs were executed, then, as a fallback, we deterministically pick a two-qubit gate $(q_1, q_2) \in \text{tg}(L)$ and SWAP along the first edge on the shortest path between $\hat{p}(q_1)$ and $\hat{p}(q_2)$. We update $\hat{p}$ according to the inserted SWAPs, update $L$, and finally update $R$. This process is repeated until the input circuit is empty.

The fallback routine ensures that this circuit transformation always produces an output circuit. The value $R$ strictly decreases in every iteration until a gate can be executed unless the fallback routine is performed, in which case $R$ stays the same. On repeated calls to the fallback routine, the same two-qubit gate is picked deterministically until it is executed. This happens within $\text{diam}(G) + 1$ iterations, where $\text{diam}(G)$ denotes the diameter of $G$. By induction we see that the whole circuit will be executed.

Let us analyze the time complexity of this circuit transformation. We ignore the initial placement since it is insignificant for large circuits. A gate from $L$ is executed in at most $\text{diam}(G)$ iterations, where $\text{diam}(G)$ is the diameter of $G$. In every iteration, $O(|E|)$ edges are checked to determine gates that can be executed and SWAPs that will decrease $R$. Therefore, the total time complexity is $O(|C||E|\text{diam}(G))$, where $|C|$ denotes the size of circuit $C$. There is a tighter bound in terms of output circuit $C'$ since every iteration creates a layer in the transformed circuit, the complexity is $O(\text{depth}(C')|E|)$, where $\text{depth}(C')$ denotes the circuit depth of $C'$.

### 2.2.3 Constructing Architecture-aware Circuit Transformations

We now present our construction for a general circuit transformation and make some definitions more precise. Let a *permuter* (Section 3) be a subroutine that, given $\pi : V \rightharpoonup V$, outputs a sequence of transpositions that implements $\pi$ while respecting the architecture constraints. Let a *mapper* (Section 4) be a subroutine that, given $\hat{p}$, a permuter, and a quantum circuit, computes a new placement of qubits, $p : Q \rightharpoonup V$, such that some gates of the input circuit can be executed.

Initialize $\hat{p}$ in the same way as the greedy swap circuit transformation. We repeat the following steps until the entire circuit has been transformed:

1. Use the given mapper to find a placement, $p : Q \rightharpoonup V$, for the remaining input circuit;

2. Let "$\circ$" denote partial function composition, i.e., given $g : X \rightharpoonup Y$ and $f : Y \rightharpoonup Z$,

$$(f \circ g)(x) := f(g(x)), \quad \text{for } x \in \text{dom}(g) \text{ and } g(x) \in \text{dom}(f). \tag{6}$$

We use the permuter to find transpositions implementing $p \circ \hat{p}^{-1} : V \rightharpoonup V$ and replace the transpositions with SWAP gates to construct a permutation circuit to execute. We also update $\hat{p}$ to reflect the new placement of qubits after running the permutation circuit.

3. Execute all gates in $L$ that can be executed in accordance with $\hat{p}$, remove these gates from the input circuit, and recompute $L$.

---

[2]This is equivalent to runnning the greedy depth mapper (Section 4.1.1) on the input circuit with only two-qubit gates, an arbitrary $\hat{p}$, and free permutations of qubits. In other words, the greedy depth mapper will pick a placement of qubits on the architecture unconstrained by movement of qubits, since this is the initial placement.

We note that the permuter used by the circuit transformation can be different from the one used by the mapper. This can, for example, be useful if the permuter is randomized and can be run multiple times in an attempt to obtain a better result. The number of such trials can be set much higher for the circuit transformation since only the permutation circuit for $p \circ \hat{p}^{-1}$ needs to be computed in every iteration. We make use of this flexibility in our implementation (Section 5).

Let us analyze the time complexity of this circuit transformation. We again ignore the time complexity of computing the initial placement. Let $t_m$ be an upper bound on the time complexity of the mapper, and let $t_p$ be an upper bound on the time complexity of the permuter. Computing $p \circ \hat{p}^{-1}$ takes time $O(|V|)$. The number of transpositions produced by the permuter is at most $t_p$, so executing the associated SWAPs takes time $O(t_p)$. Only one gate from $L$ may be executed every iteration so we upper bound the number of iterations by $|C|$. We find a time complexity of $O(|C|(t_m + |V| + t_p))$. Clearly, if $t_p, t_m \in \text{poly}(|C|, |V|)$ then our circuit transformation is also poly-time as desired.

## 3 Partial Permutations via Transpositions

In this section we provide routing algorithms for implementing partial permutations via transpositions constrained to edges of a graph. We call such algorithms *permuters*. The ROUTING VIA MATCHINGS and TOKEN SWAPPING problems capture exactly our optimization goals of implementing a permutation of qubits on a quantum architecture while minimizing the circuit depth and size, respectively.

### 3.1 Partial Routing Via Matchings

The framework of ROUTING VIA MATCHINGS captures how to permute qubits on a graph using a circuit of the smallest possible depth [2]. We first define a generalization of ROUTING VIA MATCHINGS that allows for partial permutations and then provide permuters for implementing partial permutations for some architectures of interest.

**Definition 3.1** (PARTIAL ROUTING VIA MATCHINGS). PARTIAL ROUTING VIA MATCHINGS is the following optimization problem. Given a simple graph $G = (V, E)$ and partial permutation $\pi \colon V \rightharpoonup V$, the objective is to find the smallest $k \in \mathbb{N}$ such that there exist matchings $M_1, \ldots, M_k \subseteq E$ on $G$, where each matching induces a permutation as a product of disjoint transpositions

$$\pi_{M_i} = \prod_{(v,u) \in M_i} (v\ u), \tag{7}$$

such that

$$\hat{\pi} = \prod_{i=1}^{k} \pi_{M_i} \tag{8}$$

is a completion of $\pi$.

ROUTING VIA MATCHINGS is the special case of PARTIAL ROUTING VIA MATCHINGS where $\pi$ is constrained to be a (total) permutation. The *partial routing number* of $\pi \colon V \rightharpoonup V$ on $G$ is $\text{rt}(G, \pi) \coloneqq k$, where $k$ obtains the minimum in Definition 3.1. The *routing number* [2] is the special case of the partial routing number where $\pi$ is total. In this paper, we simply refer to the partial routing number as the routing number. The routing number of $G$ is defined as

$$\text{rt}(G) \coloneqq \max_{\sigma \in \text{Sym}(V)} \text{rt}(G, \sigma), \tag{9}$$

where we maximize over all permutations $\sigma : V \leftrightarrow V$ (here $\text{Sym}(V)$ denotes the group of such permutations). Note that we only optimize over permutations, since for any $\pi : V \rightharpoonup V$,

$$\text{rt}(G, \pi) = \min_{\hat{\pi}} \text{rt}(G, \hat{\pi}), \tag{10}$$

where we minimize over all completions $\hat{\pi}$ of $\pi$.

An alternate way to interpret (PARTIAL) ROUTING VIA MATCHINGS is to assign *tokens* to all $v \in \text{dom}(\pi)$ and destinations $\pi(v)$ for the tokens. A token can only by moved through an exchange of tokens between adjacent vertices. The goal is to move all tokens to their destinations in as few matchings (specifying exchange locations) as possible. If a vertex does not hold a token at the time of an exchange with a neighbor, as can be the case in PARTIAL ROUTING VIA MATCHINGS, then after the exchange the neighbor will not hold a token.

### 3.1.1 Complete Graph

We give a simple construction of a permuter for the $n$-vertex complete graph, $K_n = (V, E)$. Given $\pi : V \rightharpoonup V$, do the following. If

$$|\text{dom}(\pi) \cup \text{image}(\pi)| = 2|\text{dom}(\pi)|, \tag{11}$$

all mappings are disjoint, so we return

$$\{(v, \pi(v)) \mid v \in \text{dom}(\pi)\} \tag{12}$$

as a single matching that implements $\pi$. Otherwise, we construct an arbitrary completion $\hat{\pi}$ of $\pi$ and run the standard algorithm for ROUTING VIA MATCHINGS for complete graphs on $\hat{\pi}$ [2]. This trivially achieves the same $\text{rt}(K_n) \leq 2$ bound for all $\pi$, but will obtain $\text{rt}(K_n, \pi) = 1$ for all $\pi$ with disjoint domain and image.

The time complexity of the ROUTING VIA MATCHINGS algorithm for $K_n$ is $O(n)$ [2]. The other operations described above also take time $O(n)$, so we get a time complexity of $O(n)$ for the complete graph permuter.

### 3.1.2 Path Graph

We construct a permuter for the $n$-vertex path graph, $P_n = (V, E)$, by first giving a completion and then using the standard complete permuter for paths [2]. Different completions achieve different routing numbers. We give a heuristic for constructing a completion that seems to result in a low routing number in practice.

We are given $\pi : V \rightharpoonup V$ and construct a completion $\hat{\pi}$ of $\pi$ as follows: Let $V \cong [n]$, ordered from one end of the path to the other (picking ends arbitrarily). Iterate through $i \in V$ in ascending order, setting

$$\hat{\pi}(i) = \begin{cases} \pi(i) & \text{if } i \in \text{dom}(\pi), \\ \min\left(V \setminus \text{image}(\hat{\pi})\right) & \text{otherwise.} \end{cases} \tag{13}$$

It can easily be seen that $\hat{\pi}$ is a completion of $\pi$. We have $\text{rt}(P_n, \pi) \leq \text{rt}(P_n, \hat{\pi}) \leq n$ by the standard path routing algorithm [2]. It remains open whether a tighter bound can be proven as a function of some parameters of $\pi$.

Constructing the completion takes time $O(|V|)$. The total complexity for running the path permuter is $O(|V|^2)$, where the time complexity of the ROUTING VIA MATCHINGS algorithm [2] dominates the construction of $\hat{\pi}$.

### 3.1.3 Hierarchical Product

The *generalized hierarchical product* (henceforth *hierarchical product*) of graphs [6] produces various subgraphs of the Cartesian product of graphs that include natural models of quantum computer architectures [5].

**Definition 3.2** (Hierarchical Product [6]). For $j \in \{1, 2\}$, let $G_j = (V_j, E_j)$ be a graph with $n_j := |V_j|$ vertices and adjacency matrix $A_j \in \mathcal{M}_{n_j}$, where $\mathcal{M}_k$ is the set of $k \times k$ boolean matrices, for $k \in \mathbb{N}$. Then the *hierarchical product* $\Pi_{\mathbf{v}}(G_1, G_2)$, for $\mathbf{v} \in \{0, 1\}^{n_2}$, has vertex set $V_1 \times V_2$ and adjacency matrix

$$A_1 \otimes \mathrm{diag}(\mathbf{v}) + \mathbb{1}_{n_1} \otimes A_2 \,,$$

where $\mathbb{1}_{n_1} \in M_{n_1}$ is the $n_1 \times n_1$ identity matrix, $M_1 \otimes M_2 \in \mathcal{M}_{n_1 n_2}$ is the Kronecker product of $M_1 \in \mathcal{M}_{n_1}$ and $M_2 \in \mathcal{M}_{n_2}$, and $\mathrm{diag}(\mathbf{v}) \in M_{n_2}$ is the diagonal matrix with the entries of $\mathbf{v}$ on the diagonal.

Intuitively, this graph consists of $n_1$ copies of $G_2$, where the $j$th vertices in all copies of $G_2$ are connected by a copy of $G_1$ if $\mathbf{v}_j = 1$. We restrict ourselves to connected simple graphs, so $A_1$ and $A_2$ are symmetric 0–1 matrices and $\mathbf{v}$ is nonzero. An example of the hierarchical product of two path graphs is

$$\Pi_{[1\ 0\ 1]}(P_2, P_3) = \Pi_{[1\ 0\ 1]}\left( \begin{matrix} 2 \\ | \\ 1 \end{matrix}, \ 1-2-3 \right) = \begin{matrix} 2,1 - 2,2 - 2,3 \\ |\quad\quad\quad\quad| \\ 1,1 - 1,2 - 1,3 \end{matrix} \tag{14}$$

The Cartesian product is $\Pi_{\mathbf{1}}$, where $\mathbf{1} := [1 \ldots 1]$ (see Section 3.1.5). Furthermore, $\Pi_{\mathbf{e}_1}$ is the standard hierarchical product, and $\Pi_{\mathbf{e}_i}$ is the rooted product of graphs, rooted at the $i$th vertex of $G_2$.

We define the vertex-induced subgraph of any graph $G = (V, E)$ for vertex set $U \subseteq V$ as

$$G[U] := (U, E \cap (U \times U)) \,. \tag{15}$$

Now, let $G = (V, E) = \Pi_{\mathbf{v}}(G_1, G_2)$ and denote the vertices of $G$ by $v = (v_1, v_2) \in V_1 \times V_2 = V$. We define

$$\mathcal{G}_i = (\mathcal{V}_i, \mathcal{E}_i) := G[\{i\} \times V_2] \,, \tag{16}$$

for $i \in V_1$. Note that each $\mathcal{G}_i$ is isomorphic to $G_2$, so the permuter for $G_2$ can be used for $\mathcal{G}_i$. We also define the *communicator vertices* of $\mathcal{G}_i$ as the vertices

$$\{i\} \times \{j \in V_2 \mid \mathbf{v}_j = 1\} \subseteq \mathcal{V}_i \,, \tag{17}$$

and index them in ascending order (for some ordering of $V$). Note that the $j$th communicator vertex (of any $\mathcal{G}_i$) also belongs to $G[V_1 \times \{j\}]$, which is isomorphic to $G_1$.

A useful metric is

$$\deg(\pi) := \max \bigcup_{i \in V_1} \{|\{v \in \mathrm{dom}(\pi) \cap \mathcal{V}_i \mid \pi(v) \notin \mathcal{V}_i\}|, |\{v \in \mathrm{dom}(\pi) \setminus \mathcal{V}_i \mid \pi(v) \in \mathcal{V}_i\}|\} \,, \tag{18}$$

which represents the maximum number of vertices that need to leave or enter any $\mathcal{G}_i$ to implement $\pi$.

In every iteration of the routing algorithm, we route a set $R = \{v^{(i)} \in \mathcal{V}_i \mid i \in V_1\}$ such that all $\pi(v)_1$ are distinct, for $v \in R$ and $\pi(v) = (\pi(v)_1, \pi(v)_2) \in V$. Undefined values are always considered distinct. We call such $R$ a set of *representative vertices*, and we view $v^{(i)}$ as the representative vertex of $V_i$.

```
    input : π : V₁ × V₂ ↪ V₁ × V₂; permuters on G₁ and G₂
1   Let Rᵢ, for i ∈ [deg(π)], be given by Lemma 3.3
2   for i = 1, ..., ⌈deg(π)/ham(v)⌉ :
3       foreach j ∈ V₁ :
4           on Gⱼ, for all k ∈ [ham(v)], route the (unique) vertex v ∈ R_{(i−1)·ham(v)+k} ∩ Vⱼ to the
            k-th communicator vertex of Gⱼ         // For R_ℓ with ℓ > deg(π), do nothing
5       foreach communicator vertex (v₁, v₂) of G₁ :              // All copies of G₁
6           on G[V₁ × {v₂}] = (V′, E′), route each v ∈ V′ ∩ dom(π) to (π(v)₁, v₂) ∈ V′
7   foreach i ∈ V₁ :
8       route all v ∈ dom(π) ∩ Vᵢ to π(v) within Gᵢ
9   return the transpositions that implement this routing
```

**Algorithm 3.1:** PARTIAL ROUTING VIA MATCHINGS on the hierarchical product of graphs $\Pi_{\mathbf{v}}(G_1, G_2)$. In Lines 4 and 6, *routing* means constructing a partial permutation $\sigma$ on a subgraph ($G_1$ or $G_2$), using the applicable permuter to find transpositions implementing $\sigma$, and applying those transpositions to update $\pi$ and each $R_i$.

**Lemma 3.3.** *For a graph $\Pi_{\mathbf{v}}(G_1, G_2)$, $\pi : V \hookrightarrow V$, let $d := \deg(\pi)$. We can find distinct sets of representative vertices $R_i$, for $i \in [d]$, such that*

$$\{v \in \mathrm{dom}(\pi) \mid v_1 \neq \pi(v)_1\} \subseteq \bigcup_{i \in [d]} R_i.$$

*Proof.* Let $G = (U, V, E)$ be a bipartite multi-graph, with $U = V := [n_1]$ the left and right vertex sets, and the edge multi-set

$$E = \{(v_1, \pi(v)_1) \mid v \in \mathrm{dom}(\pi)\}. \tag{19}$$

Each vertex $k \in U$ belongs to at most $d$ edges $(k, l)$, for $l \in V$ and $k \neq l$, and each vertex $l' \in V$ belongs to at most $d$ edges $(k', l')$, for $k' \in U$ and $k' \neq l'$. However, for any $k \in U$ there could be as many as $n_2$ edges $(k, k)$. For all $k \in U$ we remove as many $(k, k) \in E$ as necessary to ensure that the maximum degree of any vertex in $G$ is $d$.

We make $G$ $d$-regular by repeating the following: If $\nexists k \in U$ with $\deg(k) < d$ we are done. Otherwise, such a $k$ exists and $\exists k' \in V$ with $\deg(k') < d$, since

$$\sum_{k \in U} \deg(k) = \sum_{k' \in V} \deg(k'). \tag{20}$$

It follows that there exist vertices $u \in \mathcal{V}_k \setminus \mathrm{dom}(\pi)$ and $v \in \mathcal{V}_{k'} \setminus \mathrm{image}(\pi)$. For the purposes of this proof, we set $\pi(u) = v$, effectively adding an edge $(k, k')$ to $E$.

Now we have modified $\pi$ so that $G$ is $d$-regular. By Hall's marriage theorem, there exists a perfect matching in $G$, and removing it results in a $(d - 1)$-regular graph. We iterate this to find $d$ distinct perfect matchings in $G$. Each edge $(k, k') \in E$ corresponds to some $v \in \mathcal{V}_k$ and $u \in \mathcal{V}_{k'}$, with $\pi(v) = u$. Therefore, each perfect matching corresponds to a set of representative vertices, $R_i$. Since all perfect matchings are distinct, and all $e \in E$ are covered by some matching, the Lemma follows. □

Algorithm 3.1 specifies a permuter for the hierarchical product. We prove the following performance bounds for this algorithm.

**Theorem 3.4.** *For a graph $\Pi_{\mathbf{v}}(G_1, G_2)$, Algorithm 3.1 finds a sequence of transpositions that implements $\pi : V \rightharpoonup V$ certifying that*

$$\mathrm{rt}(\Pi_{\mathbf{v}}(G_1, G_2), \pi) \leq \left\lceil \frac{\deg(\pi)}{\mathrm{ham}(\mathbf{v})} \right\rceil (\mathrm{rt}(G_1) + \mathrm{rt}(G_2)) + \mathrm{rt}(G_2),$$

*where* $\mathrm{ham}(\mathbf{v})$ *is the Hamming weight of* $\mathbf{v}$*, i.e., the number of ones in* $\mathbf{v}$*.*

*Proof.* In every round of routing, we route $\mathrm{ham}(\mathbf{v})$ sets $R_i$ to their destination $\mathcal{G}_j$s, for $j \in V_1$. In each round, we route on all copies of $G_2$ in parallel and then route on all copies of $G_1$ in parallel. After routing all $R_i$ in at most $\lceil \deg(\pi) / \mathrm{ham}(\mathbf{v}) \rceil$ rounds, Lemma 3.3 ensures that only permutations local to each $\mathcal{G}_j$ remain. Finally, we route vertices to their destinations, as given by $\pi$, in each $\mathcal{G}_j$ independently using the permuter for $G_2$. □

**Corollary 3.5.**
$$\mathrm{rt}(\Pi_{\mathbf{v}}(G_1, G_2)) \leq \left\lceil \frac{n_2}{\mathrm{ham}(\mathbf{v})} \right\rceil (\mathrm{rt}(G_1) + \mathrm{rt}(G_2)) + \mathrm{rt}(G_2).$$

*Proof.* By definition (9), we maximize Theorem 3.4 over $\pi : V \leftrightarrow V$ and bound $\deg(\pi) \leq n_2$ to get the result. □

As a possible optimization, we can remove some vertices from the partial permutations in the routing steps. For each removed vertex, we must ensure that the remaining steps of the routing algorithm remain valid. Specifically, let there be a $u \in \mathcal{G}_i \cap R_k$ for $i \in V_1$ and $k \in [\deg(\pi)]$. If $u \in \mathrm{dom}(\pi)$ and $\pi(u) \in \mathcal{V}_i$, then we remove it since it does not need to be routed outside of $\mathcal{G}_i$. Otherwise, if $u \notin \mathrm{dom}(\pi)$, we remove it unless

$$\exists v \in \{R_k \cap \mathrm{dom}(\pi) \mid \pi(v) \in \mathcal{G}_i\} \tag{21}$$

since an unmapped vertex is expected at the communicator vertex in the second loop of the routing round. We apply this optimization in our implementation of the permuter for modular graphs (Section 3.1.4).

Next, we analyze the time complexity of Algorithm 3.1. Let $t_1$ and $t_2$ upper bound the time complexity of algorithms for PARTIAL ROUTING VIA MATCHINGS on $G_1$ and $G_2$, respectively. We first find $\deg(\pi)$ distinct sets of representative vertices by Lemma 3.3. The time to find one set of representative vertices is dominated by the time to find the maximum bipartite matching, $O(n_1^{2.5})$ [24]. Then, for $\lceil \deg(\pi) / \mathrm{ham}(\mathbf{v}) \rceil$ iterations, we route on all copies of $G_2$ and then $G_1$ in parallel. Overall, we get a time complexity of

$$O\left(\deg(\pi) \cdot n_1^{2.5} + \left\lceil \frac{\deg(\pi)}{\mathrm{ham}(\mathbf{v})} \right\rceil (\mathrm{ham}(\mathbf{v})t_1 + n_1 t_2) + n_1 t_2 \right). \tag{22}$$

We show a lower bound on the routing number of hierarchical products of graphs and prove that it is tight, up to constant factors.

**Theorem 3.6.** *For a graph $\Pi_{\mathbf{v}}(G_1, G_2)$ and any $\pi : V \rightharpoonup V$,*

$$2\left\lceil \frac{\deg(\pi)}{\mathrm{ham}(\mathbf{v})} \right\rceil - 1 \leq \mathrm{rt}(\Pi_{\mathbf{v}}(G_1, G_2), \pi).$$

*Proof.* Let us consider the token-based formulation of PARTIAL ROUTING VIA MATCHINGS. At most $\deg(\pi)$ tokens need to be moved out of any $\mathcal{G}_i$, for $i \in V_1$. Every matching can move at most $\mathrm{ham}(\mathbf{v})$ tokens out of their original $\mathcal{G}_i$. Once moved out, a new set of tokens must be moved onto the $\mathrm{ham}(\mathbf{v})$ communicator vertices. Therefore, it takes at least $2\lceil \deg(\pi) / \mathrm{ham}(\mathbf{v}) \rceil - 1$ matchings to move $\deg(\pi)$ tokens out of any $\mathcal{G}_i$. □

We now show that Theorem 3.6 is tight up to constant factors by considering a specific permutation on the path graph $P_{2n}$, with $n \in \mathbb{N}^+$. We have $P_{2n} = (V, E) \cong \Pi_{\mathbf{e}_1}(P_2, P_n)$ by a relabeling of vertices. We define $\pi' : V \leftrightarrow V$ as

$$\pi' := \prod_{i=0}^{n-1} (i\,(n+i)) . \tag{23}$$

Then,

$$2\left\lceil \frac{\deg(\pi')}{\operatorname{ham}(\mathbf{e}_1)} \right\rceil - 1 = 2n - 1 \leq \operatorname{rt}(\Pi_{\mathbf{e}_1}(P_2, P_n), \pi') = \operatorname{rt}(P_{2n}, \pi') \leq \operatorname{rt}(P_{2n}) \leq 2n, \tag{24}$$

where we used Section 3.1.2 for the last inequality, and $\mathbf{e}_i \in \{0,1\}^n$ is the $i$th standard basis vector. This also matches the tightest known (diameter) lower bound for $\operatorname{rt}(P_{2n})$.

### 3.1.4 Modular Graphs

Large-scale quantum computation may benefit from a modular design, with many interconnected sub-units [40, 41, 12]. As a simple model of a modular quantum processor consisting of $n_1$ modules with $n_2$ qubits each, we consider the *modular graph* $\operatorname{Mod}(n_1, n_2) := \Pi_{\mathbf{e}_1}(K_{n_1}, K_{n_2}) = (V, E)$. In this architecture, any two qubits in the same module can be directly coupled, and any two modules can be coupled through their unique communicator qubits. With one minor modification to Theorem 3.4, we get the following bounds on the routing number of the modular graph.

**Corollary 3.7.** *For $n_1, n_2 \in \mathbb{N}$ and $\pi : V \rightharpoonup V$, we have*

$$2 \deg(\pi) - 1 \leq \operatorname{rt}(\operatorname{Mod}(n_1, n_2), \pi) \leq 3 \deg(\pi) + 2 .$$

*Proof.* Directly applying Theorem 3.4 gives

$$\operatorname{rt}(\operatorname{Mod}(n_1, n_2), \pi) \leq 4 \deg(\pi) + 2 . \tag{25}$$

However, only one token needs to be routed to the communicator vertex in every round of Algorithm 3.1 and this satisfies (11). Therefore, we can route with one set of parallel transpositions, saving us one matching every round.

To show the lower bound, we apply Theorem 3.6 with $\operatorname{ham}(\mathbf{e}_1) = 1$. □

We evaluate the time complexity of this permuter using Eq. (22). Recall from Section 3.1.1 that the time complexity of the permuter is $O(n)$. Thus we have $t_1 = O(n_1)$ and $t_2 = O(n_2)$, giving an overall time complexity of $O(dn_1^{2.5} + n_1 n_2)$, where we noted that $t_2 = O(1)$ while doing the $\deg(\pi)$ rounds of routing.

### 3.1.5 Cartesian Product

The Cartesian product of graphs is a special case of the hierarchical product, namely $\Pi_{\mathbf{1}}$ for $\mathbf{1} := [1 \ldots 1]$. We refer to a copy of $G_1$ in $G_1 \times G_2$ (i.e., $G[V_1 \times \{v_2\}]$ for some $v_2 \in V_2$) as a *row* of $G_1 \times G_2$, and, vice versa, to a copy of $G_2$ as a *column*. Also, let $n_1 := |V_1|$ and $n_2 := |V_2|$. Theorem 3.4 allows us to reprove an upper bound on the routing number of a Cartesian product of graphs [2].

**Corollary 3.8.** *For any graphs $G_1 = (V_1, E_1)$ and $G_2 = (V_2, E_2)$,*

$$\operatorname{rt}(G_1 \times G_2) = \operatorname{rt}(\Pi_{\mathbf{1}}(G_1, G_2)) \leq \operatorname{rt}(G_1) + 2\operatorname{rt}(G_2) .$$

*Proof.* We fill in $\operatorname{ham}(\mathbf{v}) = n_2$ in Corollary 3.5 to get the result. □

Lemma 3.3 does not specify the order in which systems of distinct representatives are picked, but this order matters in practice. Since $\mathrm{ham}(\mathbf{v}) = n_2$, we can pick $n_2$ distinct sets of representative vertices without incurring another round of routing (in Algorithm 3.1). We propose a heuristic for picking these $n_2$ sets that seems to produce low-depth implementations of partial permutations in practice (Algorithm 3.2).

Algorithm 3.2 uses a modification of Lemma 3.3 to choose representative vertices. The proof of Lemma 3.3 can be straightforwardly extended by not initially removing edges of the form $(k, k)$ and adding edges until an $n_2$-regular bipartite multi-graph, $B$, is constructed. Thus, by Hall's marriage theorem, there exist $n_2$ distinct perfect matchings in $B$, enough for all the rows. We choose a perfect matching of minimum weight for each row with respect to a heuristic cost function $c: \mathrm{dom}(\pi) \times V_2 \to \mathbb{N}$, with the rows processed in a random order.

We add additional edges to $B$ to allow for more options to minimize the weight. We construct a bipartite multi-graph $B'$ that contains $B$, disregarding some duplicated edges. Edge duplication does not change the minimum-weight perfect matching. Instead of adding an edge for unmapped vertices as in Lemma 3.3, we add edges to all possible destination columns for each column with an unmapped vertex.

Let $\sigma: V_1 \times V_2 \rightharpoonup V_1 \times V_2$ be the partial permutation defined on Line 1 of Algorithm 3.2. The cost function depends on the current value of $\sigma$ and is defined as

$$c(v, i) := \mathrm{rt}(G_2, \pi_1) + \mathrm{rt}(G_2, \{i \mapsto \pi(v)_2\}) - \mathrm{rt}(G_2, \pi_2) - \mathrm{rt}(G_2, \{v_2 \mapsto \pi(v)_2\}), \tag{26}$$

where we define $\pi_k: V_2 \rightharpoonup V_2$ for $k \in [2]$ such that

$$\pi_1: u \mapsto \begin{cases} \sigma(v_1, u)_2 & \text{if } (v_1, u) \in \mathrm{dom}(\sigma), \\ i & \text{if } u = v_2 \end{cases} \tag{27}$$

is the partial permutation routing $v$ to row $i$ within its column, and $\pi_2: u \mapsto \sigma(v_1, u)_2$ is the current partial permutation already planned for column $v_1$. For simplicity, we assume the routing time along rows is the same in both cases, so it cancels out. To compute an upper bound on the routing number in (26) we use the given permuter for $G_2$.

To implement routing on the Cartesian product of graphs, we route $\sigma$ obtained from Algorithm 3.2 within each column independently, and proceed with Line 5 of Algorithm 3.1.

Finally, we analyze the time complexity of the permuter for Cartesian products of graphs. Assume the time complexity of computing $\mathrm{rt}(G_1, \sigma)$ and $\mathrm{rt}(G_2, \sigma')$ is upper bounded by $t_1$ and $t_2$, respectively. Computing the cost function (26) then has time complexity $O(t_2)$. In Algorithm 3.2 we construct a bipartite weighted graph with $2n_2$ vertices in time $O(n_2 n_1 t_2 + n_2^2)$. On that graph we perform a maximum weighted bipartite matching algorithm in $O(n_2^3)$ using the Hungarian algorithm [26].[3] We do this once for each row and route all vertices to their assigned rows. Then, we continue with Line 5 of Algorithm 3.1, resulting in a total time complexity for running the permuter of $O(n_1(n_2 n_1 t_2 + n_2^3) + n_2 t_1)$.

## 3.2 Partial Token Swapping

The TOKEN SWAPPING problem is similar to ROUTING VIA MATCHINGS, but minimizes the total number of transpositions instead of the depth [57]. It follows that the induced permutation circuit is optimized for circuit size. For $\epsilon > 0$, a $(1 + \epsilon)$-approximation algorithm is an algorithm that produces a solution within a factor $(1 + \epsilon)$ of optimal for all valid inputs. Here, we define a generalized version of TOKEN

---

[3] A tighter bound of $O(\sqrt{n}m \log(nC))$ is possible [19], for $n, m, C$ the number of vertices, the number of edges, and absolute maximum integer edge weight, respectively. Our edge weights can be scaled to integers that are upper bounded by $O(n_2(n_1 + n_2))$.

```
    input : π : V₁ × V₂ ⇀ V₁ × V₂, a partial permutation
1   σ ← ∅                                                          // we have σ : V₁ × V₂ ⇀ V₁ × V₂
2   r ← n₂                                                         // #remaining rows
3   foreach row i ∈_R V₂ :
4   |   E ← {(v₁, π(v)₁, c(v, i)) | v ∈ dom(π) \ dom(σ)}
    |   // Add edges for unmapped vertices
5   |   E' ← E
6   |   G = (U, V, E'), with U = V ≔ [n₁]
7   |   foreach u ∈ U with deg_E(u) < r :
8   |   |   foreach v ∈ V with deg_E(v) < r :
9   |   |   |   Add (u, v, ε) to E
10  |   Find a minimum-weight perfect matching E_match in G
11  |   V_match ← the set of vertices associated with E_match
12  |   σ ← σ + {v ↦ (v₁, i) | v ∈ V_match}                        // Recall (1)
13  |   r ← r − 1
14  return σ
```

**Algorithm 3.2:** Heuristically choosing distinct sets of representative vertices for the Cartesian product of graphs. We modify Lemma 3.3 to pick $n_2$ minimum-weight perfect matchings, with respect to the heuristic cost function $c$ (Eq. (26)). The notation $\in_R$ indicates that we select elements uniformly at random without replacement. The edges of the weighted undirected bipartite multi-graph $G$ are specified as a multi-set of triples from $V \times V \times \mathbb{R}$. We pick $\epsilon > 0$ so that zero-cost edges for mapped vertices are favored over edges for unmapped vertices.

SWAPPING that allows for partial permutations, and then give a 4-approximation algorithm for this problem on connected simple graphs that generalizes a previous 4-approximation algorithm for total permutations [39].

**Definition 3.9** (PARTIAL TOKEN SWAPPING). We define PARTIAL TOKEN SWAPPING as an optimization problem. Given are a graph $G = (V, E)$ and partial permutation $\pi : V \rightharpoonup V$. The objective is to find the smallest $k \in \mathbb{N}$ such that $\hat{\pi} = (u_1 \ v_1)(u_2 \ v_2)\ldots(u_k \ v_k)$, for $\hat{\pi}$ some completion of $\pi$ and $(u_i, v_i) \in E$ for $i \in [k]$.

Analogous to the routing number, we define the *routing size* of $\pi : V \rightharpoonup V$ on $G$, $\text{rs}(G, \pi)$, to be the minimum $k$ in Definition 3.9, and the routing size of $G$ as

$$\text{rs}(G) \coloneqq \max_{\sigma \in \text{Sym}(V)} \text{rs}(G, \sigma). \tag{28}$$

TOKEN SWAPPING is the special case of PARTIAL TOKEN SWAPPING where $\pi$ is constrained to be a total permutation. PARTIAL TOKEN SWAPPING also has an equivalent token-based formulation, similar to PARTIAL ROUTING VIA MATCHINGS.

The decision version of TOKEN SWAPPING was first shown to be NP-complete [39] and hard for a model of parametrized complexity, parametrized by the number of swaps $k$ [10]. Furthermore, assuming the Exponential Time Hypothesis (ETH), TOKEN SWAPPING cannot be solved in time $f(k)(|V| + |E|)^{o(k/\log k)}$ with $f$ any computable function [10].

```
    input : π : V ⇀ V
1   while π ≠ id |_{dom(π)} :
2       if there exists a happy swap chain v_1 v_2 ... v_ℓ then
3           Perform transpositions (v_1 v_2)(v_2 v_3) ... (v_{ℓ-1} v_ℓ)
4       else if ∃v ∈ dom(π), ∃u ∈ N(v) \ dom(π) : d(u, π(v)) < d(v, π(v)) then
5           Perform no-token swap (v u)                              // u has no token
6       else
7           There exists an unhappy swap; perform it
8       Update π according to the transpositions that were performed
9   return The sequence of transpositions that was performed
```

**Algorithm 3.3:** Routing tokens to their destinations while minimizing the number of transpositions. We add an extra step that performs no-token swaps to the algorithm of [39]. For $v \in V, N(v) \subseteq V$ denotes the set of neighbors of $v$. The partial permutation $\text{id}|_{\text{dom}(\pi)} : V \rightharpoonup V$ is the restriction of the identity function $\text{id} : V \leftrightarrow V$ to $\text{dom}(\pi)$ (so it is undefined outside of $\text{dom}(\pi)$).

### 3.2.1 Approximation Algorithm for Partial Token Swapping

We now describe a permuter that aims to minimize the circuit size. Miltzow et al. [39] gave a 4-approximation algorithm for TOKEN SWAPPING. Here, we generalize their results to PARTIAL TOKEN SWAPPING and prove that our generalized algorithm is also a 4-approximation algorithm. For this section, we consider the token-based formulation of PARTIAL TOKEN SWAPPING (recall the notion of tokens introduced in Section 3.1).

The main idea of Miltzow et al. is to perform SWAPs that reduce the sum of all distances of tokens to their destinations. We use the following definitions from [39]: An *unhappy swap* is "an edge swap where one of the tokens swapped is already on its target and the other token reduces its distance to its target vertex (by one)", and a *happy swap chain* is a path of $\ell + 1$ distinct vertices $v_1 v_2 \ldots v_\ell$, such that swapping all $(v_i, v_{i+1}) \in E$, for $i \in [\ell - 1]$, in increasing order strictly reduces the distances of all tokens in the chain to their destinations.

When considering a partial permutation, not all vertices have a token assigned to them. We add an extra step to the approximation algorithm for TOKEN SWAPPING to make use of this: Before considering an unhappy swap, we first try to swap a token to a tokenless neighbor if it brings the token closer to its destination. We call this a *no-token swap*.

The approximation algorithm for PARTIAL TOKEN SWAPPING is specified in full in Algorithm 3.3.

**Theorem 3.10.** *Given a simple connected graph $G = (V, E)$ and $\pi : V \rightharpoonup V$, Algorithm 3.3 uses at most $4 \cdot \text{rs}(G, \pi)$ transpositions.*

*Proof.* The proof is very similar to [39, Theorem 7] with some minor modifications to account for no-token swaps. Let

$$S := \sum_{v \in \text{dom}(\pi)} d(v, \pi(v)). \tag{29}$$

We know that $\text{rs}(G, \pi) \geq S/2$ since each swap can only reduce $S$ by two. A no-token swap reduces $S$ by one. A happy swap chain of length $\ell$ reduces $S$ by $\ell + 1$. As such, over the course of the algorithm,

$$\#(\text{happy swaps}) + \#(\text{no-token swaps}) \leq S. \tag{30}$$

For an unhappy swap, the token that is swapped away from its destination must next be involved in a happy swap or a no-token swap, so

$$\#(\text{unhappy swaps}) \leq \#(\text{happy swaps}) + \#(\text{no-token swaps}). \tag{31}$$

Overall, we have

$$\#(\text{unhappy swaps}) + \#(\text{happy swaps}) + \#(\text{no-token swaps})$$
$$\leq 2\#(\text{happy swaps}) + 2\#(\text{no-token swaps})$$
$$\leq 2S \leq 4 \cdot \text{rs}(G, \pi). \qquad \square$$

Miltzow et al. further showed that their algorithm for total permutations gives a 2-approximation algorithm when the graph is a tree. We now give an example showing that this is not the case for our modified algorithm when the permutation is partial. Consider the path graph $P_n$, for $n > 2$, and a partial permutation

$$\pi := \{i \mapsto i+1 \mid i \in [n-2]\} \cup \{n \mapsto 1\}. \tag{32}$$

Trivially, the shortest product of transpositions implementing $\pi$ is

$$\prod_{i=0}^{n-2}((n-i)\,(n-1-i)) \tag{33}$$

of length $n - 1$. However, the algorithm selects no-token swaps arbitrarily. In the worst case, it could select the sequence of transpositions

$$\left[\prod_{i=0}^{n-3}((n-2-i)\,(n-1-i))\right] \cdot \left[\prod_{i=0}^{n-2}((n-i)\,(n-1-i))\right] \cdot \left[\prod_{i=0}^{n-3}((2+i)\,(3+i))\right] \tag{34}$$

of length $3n - 5$. Therefore, in the limit we get an approximation ratio of $\lim_{n \to \infty}(3n - 5)/(n - 1) = 3$.

While (32) is only undefined on one input, we can modify $\pi$ by removing $k = o(n)$ entries to make it harder to find an appropriate completion, since there are $(k + 1)!$ possibilities. Then the algorithm still asymptotically achieves an approximation ratio of $\lim_{n \to \infty}(3n - 5 - 2k)/(n - 1) = 3$.

Of course, it is still possible that the algorithm could achieve better than a 4-approximation. We leave the best approximation ratio of our PARTIAL TOKEN SWAPPING algorithm (on trees and in general) as an open question.

Finally, we determine the time complexity of this permuter. Computing an all-to-all distance matrix takes time $\Theta(|V|^3)$ using the Floyd-Warshall algorithm [17], but this cost needs only to be incurred once for a graph so we do not include it. A happy or unhappy swap can be found in time $O(|E|)$ by finding cycles in an auxiliary directed graph [39]. Similarly, finding no-token swaps has time complexity $O(|E|)$. Therefore, we get a total time complexity of $O(S|E|) \leq O(|V|^2|E|)$.

## 4 Placing Qubits on the Architecture

A *mapping* algorithm (or *mapper*) finds an assignment of circuit qubits to architecture vertices such that gates can be executed efficiently. We specify mappers in terms of the routing number $\text{rt}(G, \pi)$ (Eq. (9)) and the routing size $\text{rs}(G, \pi)$ (Eq. (28)), where $G = (V, E)$ is the architecture graph and $\pi: V \rightharpoonup V$. In practice, we replace these quantities with the upper bounds that result from applying our permuters.

Mappers construct *placements* of circuit qubits onto qubits of the architecture. A placement is a bijective partial function $p: Q \rightharpoonup V$. A mapper has access to the *current placement* $\hat{p}: Q \to V$ provided by the circuit transformation. Given a placement $p$ and the current placement $\hat{p}$, we can compute a partial permutation $p \circ \hat{p}^{-1} : V \rightharpoonup V$ that implements $p$. All our mappers construct a placement $p$ that is initially undefined everywhere and modify it until finished.

In the remainder of this section, we describe several specific mappers that we implement and evaluate. We describe mappers optimizing for circuit depth in Section 4.1 and for circuit size in Section 4.2. We also give an upper bound on the time complexity of the mappers as a function of the time complexity of the permuter, $t_p$.

## 4.1 Circuit Depth Mappers

In this section we discuss mappers that attempt to minimize the transformed circuit depth. Let $L$ be the first layer of gates of the input circuit, and let $M$ be a maximum matching in the architecture graph.

### 4.1.1 Greedy Depth Mapper

The *greedy depth mapper* iteratively places the highest-cost gate at its lowest-cost location, where cost is measured in terms of the routing number to achieve the placement. More precisely, we initialize the set of used vertices $U := \emptyset$ and find a placement $p' := \{q_1 \mapsto v_1, q_2 \mapsto v_2\}$ that attains the optimum

$$\max_{(q_1,q_2)\in \text{tg}(L)} \min_{(v_1,v_2)\in M} \text{rt}\bigl(G, (p \cup \{q_1 \mapsto v_1, q_2 \mapsto v_2\}) \circ \hat{p}^{-1}\bigr), \tag{35}$$

where we consider both orderings of edges from $M$, $(v, u), (u, v) \in M$, since edges are undirected. Then, we update $U \leftarrow U \cup \text{dom}(p')$ and recompute $M$ for the graph $G[V \setminus U]$ (recall (15)); we remove the gate associated to $(q_1, q_2)$ from $L$; we set $p \leftarrow p \cup p'$; and we iterate until $\text{tg}(L) = \emptyset$ or $M = \emptyset$. Finally, we return the placement $p$.

In this procedure, we perform at most $\min\{|L|, |M|\}$ iterations to place gates. In each iteration, we find a $p'$ according to (35) in time $O(|L||M|t_p)$. Thus, the time complexity for one call of the mapper is

$$O\Bigl(\min\{|L|, |M|\}\bigl(|L||M|t_p + \sqrt{|V|}|E|\bigr)\Bigr), \tag{36}$$

where $O(\sqrt{|V|}|E|)$ is the complexity of computing a maximum matching [38].

### 4.1.2 Incremental Depth Mapper

Instead of trying to place (almost) all gates in $L$, the *incremental depth mapper* guarantees placement of only the lowest-cost gate, as given by the routing number, and incrementally improves the situation for the other gates. Specifically, we first find a placement $p_{\min} := \{q_1 \mapsto v_1, q_2 \mapsto v_2\}$ that attains the optimum

$$c'_{\min} := \min_{(q_1,q_2)\in \text{tg}(L)} \min_{(v_1,v_2)\in E} \text{rt}\bigl(G, (p \cup \{q_1 \mapsto v_1, q_2 \mapsto v_2\}) \circ \hat{p}^{-1}\bigr), \tag{37}$$

where we consider both orderings of $E$, $(u, v), (v, u) \in E$. We set $p \leftarrow p_{\min}$ and define $U := \{u, v\}$. Let $c_{\min} := \max\{c'_{\min}, 1\}$.

We find a placement for the remaining two-qubit gates that (individually) does not exceed $c_{\min}$. We iterate in arbitrary order over $(q_1, q_2) \in \text{tg}(L)$ and do the following: For $i \in [2]$, we construct a set of eligible vertices

$$U_i := \bigl\{v \in V \setminus U \mid \text{rt}\bigl(G, (p \cup \{q_i \mapsto v\}) \circ \hat{p}^{-1}\bigr) \leq c_{\min}\bigr\}. \tag{38}$$

Now we try to find $v_1^* \neq v_2^*$ as

$$(v_1^*, v_2^*) := \arg\min_{(v_1,v_2)\in U_1 \times U_2} d(v_1, v_2). \tag{39}$$

If such $(v_1^*, v_2^*)$ does not exist, we do not include $q_1$ and $q_2$ in $p$; otherwise, we set $p \leftarrow p \cup \{q_1 \mapsto v_1^*, q_2 \mapsto v_2^*\}$ and update $U \leftarrow U \cup \{v_1^*, v_2^*\}$. After iterating over all gates in tg($L$), we return $p$.

The time complexity of the incremental mapper is

$$O\bigl(|L|\bigl(|E|t_p + |V|t_p + |V|^2\bigr)\bigr). \tag{40}$$

This assumes we have access to the all-pairs distance matrix of the architecture graph, which can be precomputed in time $\Theta(|V|^3)$ [17] (independent of the input circuit).

## 4.2 Circuit Size Mappers

We now discuss mappers that optimize for circuit size. The behavior of such mappers is somewhat different from mappers optimizing for circuit depth. If there is any gate that can be performed without moving qubits, then there is no disadvantage to doing that immediately since it will have to be performed eventually. If there is any such gate, we simply return the empty placement. Thus we assume, for all mappers in this section, that there are no gates to be performed in-place.

### 4.2.1 Greedy Size Mapper

The *greedy size mapper* the same as the greedy depth mapper (Section 4.1.1), except that we replace rt($\cdot$) with rs($\cdot$) in (35).

### 4.2.2 Simple Size Mapper

The *simple size mapper* places only the lowest-cost gate at its lowest-cost location. More precisely, we find a placement $p \coloneqq \{q_1 \mapsto v_1, q_2 \mapsto v_2\}$ that attains the optimum

$$\min_{(q_1,q_2)\in \text{tg}(L)} \min_{(v_1,v_2)\in E} \text{rs}\bigl(G, (p \cup \{q_1 \mapsto v_1, q_2 \mapsto v_2\}) \circ \hat{p}^{-1}\bigr) \tag{41}$$

where we consider all orderings of the edges of $E$, and return $p$. Note that we have replaced rt($\cdot$) with rs($\cdot$) in (37). The time complexity of the simple size mapper is $O(|L||E|t_p)$.

### 4.2.3 Extension Size Mapper

The *extension size mapper* first finds an initial placement $p$ using (41). Let $c'_{\min}$ be the value attained at the optimum for (41). After finding the initial placement, we try to only place another gate if it is cheaper to place now rather than in a later call to the mapper.

Specifically, for the current $p$ and $\hat{p}$, we define $\hat{p}' : Q \to V$ as the placement after performing the permutation circuit constructed from transpositions achieving rs($G, p \circ \hat{p}^{-1}$). Let $U \coloneqq \emptyset$. Now we define a heuristic for the number of saved transpositions, $s : Q \times Q \to \mathbb{N}$, as

$$\begin{aligned}
s(q_1, q_2) \coloneqq &\,\text{rs}\bigl(G, p \circ \hat{p}^{-1}\bigr) + \min_{(v_1,v_2)\in E} \text{rs}\bigl(G, \{q_1 \mapsto v_1, q_2 \mapsto v_2\} \circ (\hat{p}')^{-1}\bigr) \\
&- \min_{(u_1,u_2)\in E'} \text{rs}\bigl(G, (p \cup \{q_1 \mapsto u_1, q_2 \mapsto u_2\}) \circ \hat{p}^{-1}\bigr),
\end{aligned} \tag{42}$$

where $E'$ is the edge set of $G[V \setminus U]$ and we consider all orderings of the edges of $E$ and $E'$.

The extension size mapper iterates the following. We find the gate $(q_1^*, q_2^*) \in \text{tg}(L)$ attaining

$$s_{\max} \coloneqq \max_{(q_1,q_2)\in \text{tg}(L)} s(q_1, q_2), \tag{43}$$

and let $(u_1^*, u_2^*) \in E'$ be the edge attaining $s_{\max}$ as given by (42). If $s_{\max} \geq 0$, we set $p \leftarrow p \cup \{q_1^* \mapsto u_1^*, q_2^* \mapsto u_2^*\}$, remove the gate $(q_1^*, q_2^*)$ from $L$, update $U \leftarrow U \cup \{v_1^*, v_2^*\}$, and iterate; otherwise, we stop and return $p$.

Calculating $s(q_1, q_2)$ for any $q_1, q_2 \in Q$ takes time $O(|E|t_p)$. Therefore, the total time complexity of the extension size mapper is
$$O(|L|^2 |E| t_p). \tag{44}$$

### 4.2.4 Qiskit-based Mapper

Finally, we implement a mapper that is based on Qiskit's circuit transformation (described in Section 2.2.1). Since this is a mapper, we only execute one iteration of the circuit transformation: for the first layer $L$. We also do not modify the output circuit, but instead return the final $\hat{p}$ that would be induced by executing all SWAPs found during the mapping process.

We make three changes to Qiskit's circuit transformation. The first is that when minimizing $S$, instead of choosing a maximal set of SWAPs in every iteration, we choose only one SWAP along an edge $e \in E$ that minimizes $S$. The second is that the upper bound on the number of iterations is raised to $|V|^2$, since we only apply one SWAP per iteration. Thirdly, if no trial is successful, we fall back to the simple size mapper (Section 4.2.2) and return the placement it finds, which places only one gate in this iteration.

We now give the time complexity of the Qiskit mapper. First, we compute an all-to-all distance matrix in time $\Theta(|V|^3)$ [17], which we ignore since it is a one-time cost dependent only on the architecture. Each of the $O(|V|^2)$ iterations has a time complexity of $O(|E||L|)$. Thus, the Qiskit mapper has time complexity $O(|V|^2 |E||L|)$.

## 5 Results

We implement the circuit transformation introduced in Section 2.2.3 with a variety of mappers and appropriate permuters. We also implement the greedy swap transformation described in Section 2.2.2. We check the validity of our implementations by testing closeness in fidelity of the original output state and that of the transformed circuit for random input states of 11 qubits on random circuits [48] (described in the next section).

### 5.1 Evaluation Criteria

When testing the performance of these circuit transformations, each is allocated at most 8GB of RAM and 2 days to transform all circuits of a data point. For each data point we transform 10 random circuits and 1 QSP circuit. We consider a 2-day runtime acceptable, given that classical computational resources are plentiful compared to quantum ones. We generate the data on a heterogeneous cluster with Intel Opteron 2354 and Intel Xeon X5560 processors.

The Cartesian permuter (Section 3.1.5), the general size permuter (Section 3.2.1), and Qiskit's circuit transformation (Section 2.2.1) are randomized. We run multiple trials of these permuters and take the best result. Most of the time, trials produce equally good permutation circuits, although occasionally they deviate by a few SWAP gates. Our mappers run permuters $O(|L||E|)$ times, so we do only 4 trials to quickly remove any bad outliers. In contrast, our circuit transformation only directly runs a permuter once per layer of gates, so in this case we perform a slower 100 trials in an attempt to save a few SWAPs. We leave the number of trials for Qiskit's circuit transformation at its default of 40.

We test the performance of circuit transformations for the grid, $P_{n_1} \times P_{n_2}$, using the permuter from Section 3.1.5 and the modular architecture, $\text{Mod}(n_1, n_2)$, with the permuter from Section 3.1.4, for $n_1, n_2 \in \mathbb{N}$. For an $N$-qubit circuit, we set $n_1 = n_2 = \lceil \sqrt{N} \rceil$ so that there are enough qubits in the architecture to contain the circuit. By Corollary 3.8, we know that taking $n_1 = n_2$ minimizes the routing time for our routing strategy among all grids with the same number of qubits. It is less clear how to balance parameters for the modular architecture since Corollary 3.7 does not depend on $n_1$ and $n_2$. For $n_1 \ll n_2$ or $n_2 \ll n_1$, less movement of qubits is needed, since many qubits are adjacent to one another. Thus, we take $n_1 = n_2$ in an attempt to consider a hard case. For some values of $N$, it may also be possible to find parameters $n_1' \neq n_2'$ such that $N \leq n_1' n_2' < \lceil \sqrt{N} \rceil^2 = n_1 n_2$, requiring fewer qubits. However, this introduces unwanted size-dependent behavior in our results when $|n_1' - n_2'| \gg 0$ for one circuit size and $n_1' \approx n_2'$ for the next, so we find it preferable to fix $n_1 = n_2$.

We compare the transformed circuits in terms of their *weighted depth* and *weighted size*. For both trapped-ion and superconducting qubits, two-qubit gates typically have longer execution times and lower fidelities than single-qubit gates [32]. Even among two-qubit gates there is a difference between execution times. Assuming fast local unitaries, the SWAP gate has 1–3 times the interaction cost of a CNOT depending on the physical interactions used to realize the gates [53]. For simplicity, we assign unit cost for one-qubit gates, cost 10 for CNOT, and cost 30 for SWAP. We define the weighted size of a circuit as the sum of all gate weights and the weighted depth of a circuit as the maximum-weight path in the DAG of the circuit, where the weight of a path is the sum of the weights of the gates along it.

We consider two circuit families: random circuits and quantum signal processing (QSP) circuits [33]. Random circuits have been proposed for quantum computational supremacy experiments on near-term quantum devices [9, 11]. Such proposals typically construct random circuits so that architecture constraints are automatically obeyed. For our purposes, random circuits provide a class of examples with little structure for circuit transformations to exploit, so we expect them to represent a hard case with large overhead. We generate a fixed set of 10 random circuits for various qubit counts. We set the number of circuit layers to 20. For each layer, we bin the qubits into pairs uniformly at random and assign each pair of qubits a Haar-random unitary from SU(4). Finally, we decompose each unitary into the smallest possible number of CNOT + SU(2) gates [51]. This random circuit generator is provided by Qiskit [8].

We consider QSP circuits for Hamiltonian simulation as an example of a realistic quantum algorithm. We use the unoptimized circuits provided in [14], decomposed into $Z$ rotations, CNOT gates, and single-qubit Clifford gates. The QSP algorithm requires precise angles that turn out to be expensive to compute. Therefore, [14] uses randomized angles instead, giving a circuit that does not correctly implement the Hamiltonian simulation. Nevertheless, the circuit corresponds to an accurate implementation of QSP, up to rotation angles, and can be used for benchmarking resources. Furthermore, the circuit transformations we construct are unaffected by those angles. We only consider one pair of *phased iterates* of the QSP algorithm ($V_{\varphi_i + \pi}^\dagger V_{\varphi_{i-1}}$ as in [14, Eq. 31]). A full QSP circuit for the architecture can be constructed by iterating the mapped circuit of such phased iterates, a permutation circuit between iterations, state preparation, and state unpreparation. The cost of the transformed phased iterates dominates all other costs of the construction, so the total cost can be estimated by taking our result times the number of iterations.

The circuit transformations from Section 2.2.3 are constructed from a permuter and a mapper. We denote such circuit transformations by tf : {d,s} × $\mathcal{M}$, where $\mathcal{M}$ is the set of all mappers (see Table 2), "d" denotes an appropriate depth permuter (Section 3.1), and "s" denotes the general size permuter (Section 3.2.1). For example, by tf(d,greedy depth) we denote a circuit transformation with a depth permuter for the architecture and the greedy depth mapper (Section 4.1.1).

| Abbreviation | Mapper name | Section |
|---|---|---|
| greedy depth | greedy depth mapper | 4.1.1 |
| incremental | incremental depth mapper | 4.1.2 |
| greedy size | greedy size mapper | 4.2.1 |
| simple | simple size mapper | 4.2.2 |
| extend | extension size mapper | 4.2.3 |
| qiskit | qiskit-based mapper | 4.2.4 |

**Table 2:** The abbreviated names of the set of mappers $\mathcal{M}$ used to construct circuit transformations tf : $\{d,s\} \times \mathcal{M}$.

## 5.2 Numerical Results

Figure 1 plots our results. We first consider the random circuit results. For the grid, we find that tf(d,incremental) shows much slower growth of weighted depth than circuit transformations that do not use depth-optimized permuters (Section 3.1.5). We also note that tf(d,qiskit) performs much better than Qiskit's circuit transformation (Section 2.2.1), suggesting that depth-optimized permuters can offer a significant advantage. On the modular graph, Qiskit's circuit transformation is much better at minimizing the weighted depth, but tf(d,qiskit) starts closing the gap for larger sizes. Unfortunately, we do not know if tf(d,qiskit) performs better at larger sizes because Qiskit's circuit transformation is not fast enough to generate the relevant data. Up to 100 qubits tf(s,qiskit) achieves the best weighted size on grid architectures, and tf(s,simple) does best on modular architectures up to 121 qubits. For all sizes the greedy swap circuit transformation (Section 2.2.2) performs as one of the best at optimizing for weighted circuit size. The greedy swap circuit transformation is also able transform larger circuits within the time limit as expected from its lower time complexity.

For larger QSP circuits, the greedy circuit transformation (Section 2.2.2) is the clear winner in both weighted depth and weighted size, suggesting that it may be a good approach for practical quantum circuits. Surprisingly, tf(s,qiskit) also performs fairly well at minimizing the depth despite targeting the circuit size.

## 6 Conclusion and Future Work

We have specified various ways to efficiently transform general quantum circuits to respect architecture constraints while attempting to minimize the overhead. We investigated the qubit movement subproblem and proposed PARTIAL ROUTING VIA MATCHINGS and PARTIAL TOKEN SWAPPING as models of our optimization objectives of minimizing the circuit depth and circuit size, respectively. We gave algorithms for PARTIAL ROUTING VIA MATCHINGS for the path graph, the complete graph, and for the generalized hierarchical products of graphs, and showed tighter bounds for certain partial permutations. We then gave more detailed analyses of special cases of the generalized hierarchical product that arise in proposed quantum architectures: the Cartesian product (e.g., for grid architectures) and the modular architecture. We also showed a 4-approximation algorithm for PARTIAL TOKEN SWAPPING on general graphs.

We constructed circuit transformations with a variety of heuristic qubit placement strategies (called *mappers*). A mapper attempts to find suitable qubit placements on the architecture to execute the circuit succinctly. Given a permuter subroutine, our mappers can handle any connected simple graph. We also showed how to construct a circuit transformation from a permuter and a mapper.

Finally, we tested our circuit transformations against Qiskit's circuit transformation and a basic greedy strategy with large quantum circuits on a grid or modular architecture. When optimizing for weighted circuit size, our greedy circuit transformation was one of the best in all cases, though using our circuit transformations with algorithms for PARTIAL TOKEN SWAPPING sometimes gave a slight advantage. For the weighted circuit depth, the picture was more nuanced. We found that algorithms using PARTIAL ROUTING VIA MATCHINGS for qubit movement could give good performance for random circuits, but Qiskit's circuit transformation and our greedy circuit transformation also performed well and gave the best results in some cases.

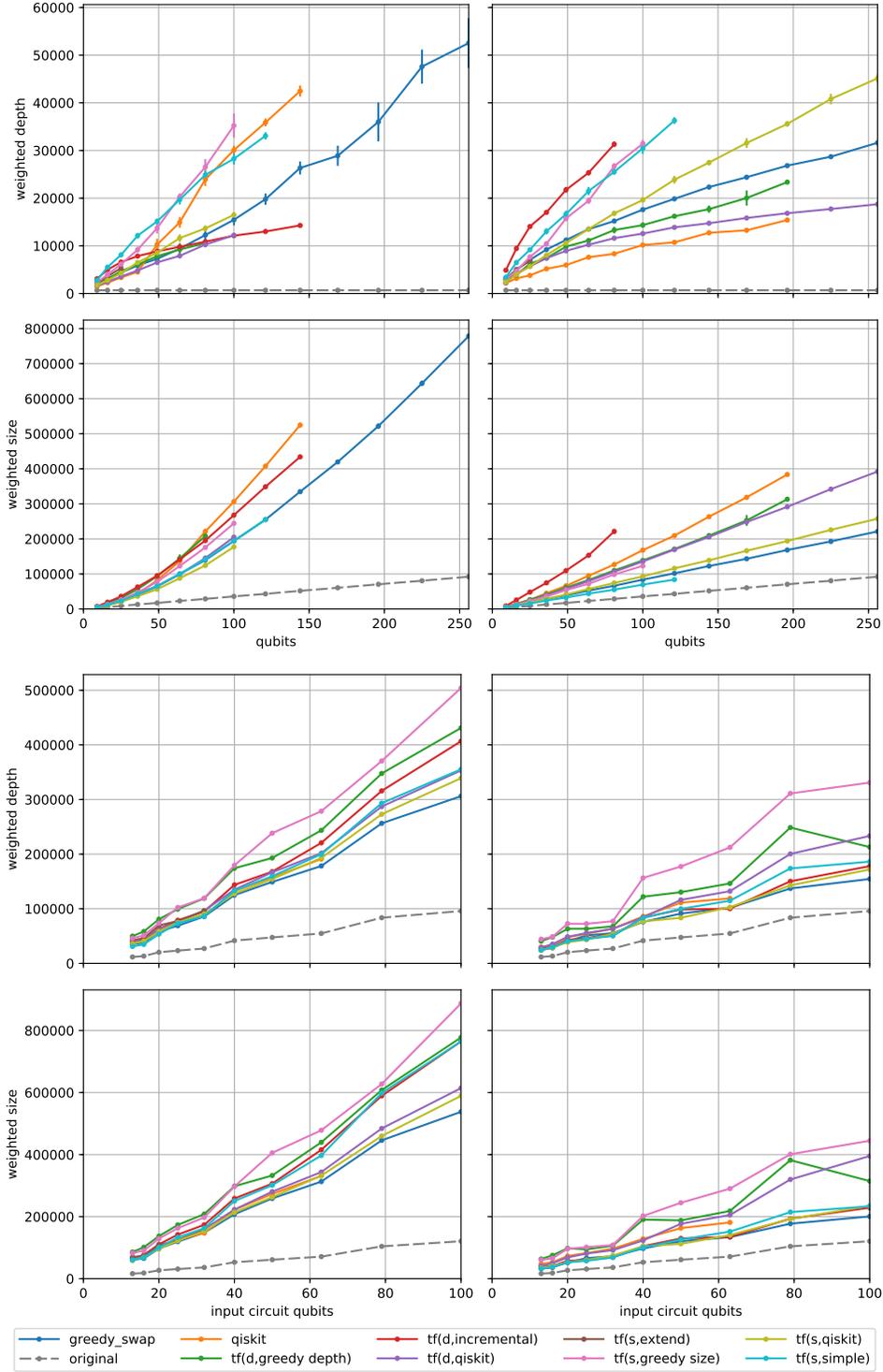

**Figure 1:** The weighted depth and weighted size for transformed random circuits (top two rows) and QSP circuits (bottom two rows) on the grid architecture (left column) and the modular architecture (right column). We generate fixed sets of 10 random circuits for increasing qubit counts and plot the mean and standard deviation for each data point. One QSP circuit is considered for each data point. The metrics for the original circuit are also given to make the overhead introduced in circuit transformations explicit; note that the original circuit does not respect the architecture constraints. The notation tf : {d,s} × $\mathcal{M}$ indicates a circuit transformation constructed from either an appropriate depth ("d") permuter or the size ("s") permuter and one of our mappers (Table 2).

We would like to better understand what circuit transformations work best for which architectures, quantum algorithms, and objective functions. We also would like to use the tools of PARTIAL ROUTING VIA MATCHINGS and PARTIAL TOKEN SWAPPING to establish bounds on the overhead of specific quantum architectures. Ideally, we could use these tools and circuit transformations to design architectures that offer good performance subject to realistic hardware constraints and to compute realistic resource estimates for implementations of quantum algorithms.

There are many ways our methods could be improved. It would be interesting to know whether one can do better than just using SWAP gates to route qubits. Our mapper algorithms may also be improved by including some form of lookahead to consider later layers of the given circuit, or by specializing mappers to particular architectures.

Modeling the architecture as a simple graph loses information about the underlying hardware. For example, in the IBM system the architecture edges have directionality indicating the control and target of CNOTs. In implementations of the modular architecture, the interconnecting links are probably much noisier and slower than local operations. In general, gate costs and times can vary significantly across a hardware implementation and sometimes even vary over time [42]. Adapting to variable costs and keeping track of operations performed asynchronously is challenging but could be worthwhile for architectures that support a mixture of fast and slow operations.

Finally, we hope that future progress on the challenges addressed in this paper will be facilitated by a suitable set of benchmarks of large quantum circuits. We publicly make available and license our source code, benchmark circuits, and results (in TSV format) [48] and encourage others to do the same.

## Acknowledgments

The authors would like to thank Aniruddha Bapat for insights in the hierarchical product of graphs and suggestions for tightening the routing lower bound on these graphs. We would also like to thank Drew Risinger for helpful formative discussions. This work was supported in part by the Army Research Office (MURI award number W911NF-16-1-0349), the Canadian Institute for Advanced Research, the National Science Foundation (grant number 1813814), and the U.S. Department of Energy, Office of Science, Office of Advanced Scientific Computing Research, Quantum Algorithms Teams and Quantum Testbed Pathfinder (award number DE-SC0019040) programs.## References

[1] M. Ajtai, J. Komlós, and E. Szemerédi. "An $O(n \log n)$ sorting network". In: *Proceedings of the fifteenth annual ACM symposium on theory of computing (STOC)*. ACM Press, 1983, pp. 1–9. ISBN: 0-89791-099-0. DOI: 10.1145/800061.808726.

[2] Noga Alon, F. R. K. Chung, and R. L. Graham. "Routing permutations on graphs via matchings". In: *SIAM journal on discrete mathematics* 7.3 (May 1994), pp. 513–530. ISSN: 0895-4801. DOI: 10.1137/s0895480192236628.

[3] Indranil Banerjee and Dana Richards. "New results on routing via matchings on graphs". In: *Fundamentals of computation theory*. Springer Berlin Heidelberg, 2017, pp. 69–81. DOI: 10.1007/978-3-662-55751-8_7.

[4] Indranil Banerjee, Dana Richards, and Igor Shinkar. "Sorting networks on restricted topologies". In: *SOFSEM 2019: theory and practice of computer science*. Springer International Publishing, 2019, pp. 54–66. DOI: 10.1007/978-3-030-10801-4_6.

[5] Aniruddha Bapat, Zachary Eldredge, James R. Garrison, Abhinav Deshpande, Frederic T. Chong, and Alexey V. Gorshkov. "Unitary entanglement construction in hierarchical networks". In: *Physical review a* 98.6 (Dec. 26, 2018). DOI: 10.1103/PhysRevA.98.062328.

[6] L. Barrière, C. Dalfó, M. A. Fiol, and M. Mitjana. "The generalized hierarchical product of graphs". In: *Discrete mathematics* 309.12 (June 2009), pp. 3871–3881. ISSN: 0012-365X. DOI: 10.1016/j.disc.2008.10.028.

[7] R. Beals, S. Brierley, O. Gray, A. W. Harrow, S. Kutin, N. Linden, D. Shepherd, and M. Stather. "Efficient distributed quantum computing". In: *Proceedings of the royal society a: mathematical, physical and engineering sciences* 469.2153 (Feb. 20, 2013). DOI: 10.1098/rspa.2012.0686.

[8] Luciano Bello, Jim Challenger, Andrew Cross, Ismael Faro, Jay Gambetta, Juan Gomez, Ali Javadi-Abhari, Paco Martin, Diego Moreda, Jesus Perez, Erick Winston, and Chris Wood. *Qiskit. An open source quantum computing framework for writing quantum experiments, programs, and applications*. IBM. 2017. URL: https://www.qiskit.org/ (visited on 08/21/2018).

[9] Sergio Boixo, Sergei V. Isakov, Vadim N. Smelyanskiy, Ryan Babbush, Nan Ding, Zhang Jiang, Michael J. Bremner, John M. Martinis, and Hartmut Neven. "Characterizing quantum supremacy in near-term devices". In: *Nature physics* 14.6 (June 2018), pp. 595–600. DOI: 10.1038/s41567-018-0124-x.

[10] Édouard Bonnet, Tillmann Miltzow, and Paweł Rzążewski. "Complexity of token swapping and its variants". In: *Algorithmica* 80.9 (Oct. 2017), pp. 2656–2682. DOI: 10.1007/s00453-017-0387-0.

[11] Adam Bouland, Bill Fefferman, Chinmay Nirkhe, and Umesh Vazirani. "On the complexity and verification of quantum random circuit sampling". In: *Nature physics* (Oct. 29, 2018). DOI: 10.1038/s41567-018-0318-2.

[12] Teresa Brecht, Wolfgang Pfaff, Chen Wang, Yiwen Chu, Luigi Frunzio, Michel H. Devoret, and Robert J. Schoelkopf. "Multilayer microwave integrated quantum circuits for scalable quantum computing". In: *Npj quantum information* 2.16002 (Feb. 23, 2016). DOI: 10.1038/npjqi.2016.2.

[13] Stephen Brierley. "Efficient implementation of quantum circuits with limited qubit interactions". In: *Quantum info. comput.* 17.13-14 (Nov. 2017), pp. 1096–1104. ISSN: 1533-7146. arXiv: 1507.04263 [quant-ph].

[14] Andrew M. Childs, Dmitri Maslov, Yunseong Nam, Neil J. Ross, and Yuan Su. "Toward the first quantum simulation with quantum speedup". In: *Proceedings of the national academy of sciences* 115.38 (Sept. 18, 2018), pp. 9456–9461. DOI: 10.1073/pnas.1801723115.

[15] Byung-Soo Choi and Rodney van Meter. "A $\Theta(\sqrt{N})$-depth quantum adder on the 2D NTC quantum computer architecture". In: *Acm journal on emerging technologies in computing systems* 8.3 (Aug. 3, 2012), 24:1–24:22. ISSN: 1550-4832. DOI: 10.1145/2287696.2287707.

[16] Byung-Soo Choi and Rodney van Meter. "On the effect of quantum interaction distance on quantum addition circuits". In: *ACM journal on emerging technologies in computing systems* 7.3 (Aug. 2011), pp. 1–17. DOI: 10.1145/2000502.2000504.

[17] Robert W. Floyd. "Algorithm 97: shortest path". In: *Communications of the ACM* 5.6 (June 1962), p. 345. DOI: 10.1145/367766.368168.

[18] Austin G. Fowler, Simon J. Devitt, and Lloyd C. L. Hollenberg. "Implementation of Shor's algorithm on a linear nearest neighbour qubit array". In: *Quantum information & computation* 4 (Feb. 25, 2004), pp. 237–251. arXiv: quant-ph/0402196v1.

[19] Harold N. Gabow and Robert E. Tarjan. "Faster scaling algorithms for network problems". In: *SIAM journal on computing* 18.5 (Oct. 1989), pp. 1013–1036. DOI: 10.1137/0218069.

[20] Google Quantum AI Lab. *A preview of Bristlecone, Google's new quantum processor*. Mar. 5, 2018. URL: https://ai.googleblog.com/2018/03/a-preview-of-bristlecone-googles-new.html (visited on 10/09/2018).

[21] Jeff Heckey, Shruti Patil, Ali JavadiAbhari, Adam Holmes, Daniel Kudrow, Kenneth R. Brown, Diana Franklin, Frederic T. Chong, and Margaret Martonosi. "Compiler management of communication and parallelism for quantum computation". In: *Proceedings of the twentieth international conference on architectural support for programming languages and operating systems (ASPLOS)*. ACM Press, 2015. DOI: 10.1145/2694344.2694357.

[22] Steven Herbert. *On the depth overhead incurred when running quantum algorithms on near-term quantum computers with limited qubit connectivity*. Sept. 25, 2018. arXiv: 1805.12570v4 [quant-ph].

[23] Yuichi Hirata, Masaki Nakanishi, Shigeru Yamashita, and Yasuhiko Nakashima. "An efficient method to convert arbitrary quantum circuits to ones on a linear nearest neighbor architecture". In: *2009 third international conference on quantum, nano and micro technologies*. IEEE, Feb. 2009. DOI: 10.1109/icqnm.2009.25.

[24] John E. Hopcroft and Richard M. Karp. "An $n^{5/2}$ algorithm for maximum matchings in bipartite graphs". In: *SIAM journal on computing* 2.4 (Dec. 1973), pp. 225–231. DOI: 10.1137/0202019.

[25] Donald E. Knuth. "Networks for sorting". In: Second. Vol. 3. The Art of Computer Programming. Addison-Wesley Professional, 1998, pp. 219–247. ISBN: 0201896850.

[26] H. W. Kuhn. "The hungarian method for the assignment problem". In: *Naval research logistics quarterly* 2.1-2 (Mar. 1955), pp. 83–97. DOI: 10.1002/nav.3800020109.

[27] Manfred Kunde. "Optimal sorting on multi-dimensionally mesh-connected computers". In: *Proceedings of the symposium on theoretical aspects of computer science (STACS)*. Vol. 247. Lecture Notes in Computer Science. Berlin, Heidelberg: Springer Berlin Heidelberg, 1987, pp. 408–419. ISBN: 978-3-540-47419-7. DOI: 10.1007/bfb0039623.

[28] Samuel A. Kutin, David Petrie Moulton, and Lawren M. Smithline. "Computation at a distance". In: *Chicago journal of theoretical computer science* 13.1 (Sept. 25, 2007), pp. 1–17. DOI: 10.4086/cjtcs.2007.001.

[29] L. Lao, B. van Wee, I. Ashraf, J. van Someren, N. Khammassi, K. Bertels, and C. G. Almudever. "Mapping of lattice surgery-based quantum circuits on surface code architectures". In: *Quantum science and technology* 4.1 (Sept. 2018), p. 015005. DOI: 10.1088/2058-9565/aadd1a.

[30] Gushu Li, Yufei Ding, and Yuan Xie. "Tackling the qubit mapping problem for nisq-era quantum devices". In: *Proceedings of the twenty-fourth international conference on architectural support for programming languages and operating systems (ASPLOS)*. Ed. by Iris Bahar, Maurice Herlihy, Emmett Witchel, and Alvin R. Lebeck. Providence, RI, USA: ACM, 2019, pp. 1001–1014. ISBN: 978-1-4503-6240-5. DOI: 10.1145/3297858.3304023.

[31] Chia-Chun Lin, Susmita Sur-Kolay, and Niraj K. Jha. "PAQCS: physical design-aware fault-tolerant quantum circuit synthesis". In: *IEEE transactions on very large scale integration (VLSI) systems* 23.7 (July 2015), pp. 1221–1234. DOI: 10.1109/tvlsi.2014.2337302.

[32] Norbert M. Linke, Dmitri Maslov, Martin Roetteler, Shantanu Debnath, Caroline Figgatt, Kevin A. Landsman, Kenneth Wright, and Christopher Monroe. "Experimental comparison of two quantum computing architectures". In: *Proceedings of the national academy of sciences* 114.13 (Mar. 28, 2017), pp. 3305–3310. DOI: 10.1073/pnas.1618020114.

[33] Guang Hao Low and Isaac L. Chuang. "Optimal Hamiltonian simulation by quantum signal processing". In: *Physical review letters* 118.1 (Jan. 2017). DOI: 10.1103/physrevlett.118.010501.

[34] Aaron Lye, Robert Wille, and Rolf Drechsler. "Determining the minimal number of SWAP gates for multi-dimensional nearest neighbor quantum circuits". In: *20th asia and south pacific design automation conference (ASP-DAC)*. IEEE, 2015. ISBN: 978-1-4799-7792-5. DOI: 10.1109/aspdac.2015.7059001.

[35] D. Maslov, S. M. Falconer, and M. Mosca. "Quantum circuit placement". In: *IEEE transactions on computer-aided design of integrated circuits and systems* 27.4 (Mar. 21, 2008), pp. 752–763. DOI: 10.1109/tcad.2008.917562.

[36] Dmitri Maslov. "Linear depth stabilizer and quantum Fourier transformation circuits with no auxiliary qubits in finite-neighbor quantum architectures". In: *Physical review a* 76.5 (Nov. 2007). DOI: 10.1103/physreva.76.052310.

[37] Tzvetan S. Metodi, Darshan D. Thaker, Andrew W. Cross, Frederic T. Chong, and Isaac L. Chuang. "Scheduling physical operations in a quantum information processor". In: *Quantum information and computation IV*. Ed. by Eric J. Donkor, Andrew R. Pirich, and Howard E. Brandt. Vol. 6244. SPIE, May 12, 2006. DOI: 10.1117/12.666419.

[38] Silvio Micali and Vijay V. Vazirani. "An $O(\sqrt{|V|}|E|)$ algorithm for finding maximum matching in general graphs". In: *21st annual symposium on foundations of computer science (sfcs)*. IEEE, Oct. 1980. DOI: 10.1109/sfcs.1980.12.

[39] Tillmann Miltzow, Lothar Narins, Yoshio Okamoto, Günter Rote, Antonis Thomas, and Takeaki Uno. "Approximation and hardness of token swapping". In: *24th annual european symposium on algorithms (ESA)*. Ed. by Piotr Sankowski and Christos Zaroliagis. Vol. 57. Leibniz International Proceedings in Informatics (LIPIcs). Dagstuhl, Germany: Schloss Dagstuhl–Leibniz-Zentrum fuer Informatik, Aug. 2016, 66:1–66:15. ISBN: 978-3-95977-015-6. DOI: 10.4230/LIPIcs.ESA.2016.66.

[40] C. Monroe and J. Kim. "Scaling the ion trap quantum processor". In: *Science* 339.6124 (Mar. 8, 2013), pp. 1164–1169. DOI: 10.1126/science.1231298.

[41] C. Monroe, R. Raussendorf, A. Ruthven, K. R. Brown, P. Maunz, L.-M. Duan, and J. Kim. "Large-scale modular quantum-computer architecture with atomic memory and photonic interconnects". In: *Physical review a* 89.2 (Feb. 2014). DOI: 10.1103/physreva.89.022317.

[42] Prakash Murali, Jonathan M. Baker, Ali Javadi-Abhari, Frederic T. Chong, and Margaret Martonosi. "Noise-adaptive compiler mappings for noisy intermediate-scale quantum computers". In: *Proceedings of the twenty-fourth international conference on architectural support for programming languages and operating systems (ASPLOS)*. Ed. by Iris Bahar, Maurice Herlihy, Emmett Witchel, and Alvin R. Lebeck. Providence, RI, USA: ACM, 2019, pp. 1015–1029. ISBN: 978-1-4503-6240-5. DOI: 10.1145/3297858.3304075.

[43] Flemming Nielson, Hanne Riis Nielson, and Chris Hankin. *Principles of program analysis*. Springer Berlin Heidelberg, 1999. DOI: 10.1007/978-3-662-03811-6.

[44] M. Pedram and A. Shafaei. "Layout optimization for quantum circuits with linear nearest neighbor architectures". In: *Ieee circuits and systems magazine* 16.2 (2016), pp. 62–74. ISSN: 1531-636X. DOI: 10.1109/MCAS.2016.2549950.

[45] Rigetti. *QPU specifications. Rigetti 16q aspen-1*. 2018. URL: https://www.rigetti.com/qpu (visited on 01/23/2019).

[46] David J. Rosenbaum. "Optimal quantum circuits for nearest-neighbor architectures". In: *8th conference on the theory of quantum computation, communication and cryptography (TQC)*. Ed. by Simone Severini and Fernando Brandao. Vol. 22. Leibniz International Proceedings in Informatics (LIPIcs). Dagstuhl, Germany: Schloss Dagstuhl–Leibniz-Zentrum fuer Informatik, 2013, pp. 294–307. ISBN: 978-3-939897-55-2. DOI: 10.4230/LIPIcs.TQC.2013.294.

[47] Mehdi Saeedi, Robert Wille, and Rolf Drechsler. "Synthesis of quantum circuits for linear nearest neighbor architectures". In: *Quantum information processing* 10.3 (June 1, 2011), pp. 355–377. ISSN: 1573-1332. DOI: 10.1007/s11128-010-0201-2.

[48] Eddie Schoute, Cem Unsal, and Andrew Childs. *Arct. Architecture-respect circuit transformations*. 2019. URL: https://gitlab.umiacs.umd.edu/amchilds/arct (visited on 02/24/2019).

[49] Alireza Shafaei, Mehdi Saeedi, and Massoud Pedram. "Qubit placement to minimize communication overhead in 2D quantum architectures". In: *19th asia and south pacific design automation conference (ASP-DAC)*. IEEE, Jan. 2014. ISBN: 978-1-4799-2816-3. DOI: 10.1109/aspdac.2014.6742940.

[50] IBM Q Team. *IBM Q Experience devices*. 2018. URL: https://quantumexperience.ng.bluemix.net/qx/devices (visited on 10/09/2018).

[51] Farrokh Vatan and Colin Williams. "Optimal quantum circuits for general two-qubit gates". In: *Physical review a* 69.3 (Mar. 22, 2004). DOI: 10.1103/physreva.69.032315.

[52] Davide Venturelli, Minh Do, Eleanor Rieffel, and Jeremy Frank. "Compiling quantum circuits to realistic hardware architectures using temporal planners". In: *Quantum science and technology* 3.2 (Feb. 2018), p. 025004. DOI: 10.1088/2058-9565/aaa331.

[53] G. Vidal, K. Hammerer, and J. I. Cirac. "Interaction cost of nonlocal gates". In: *Physical review letters* 88 (23 May 2002), p. 237902. DOI: 10.1103/PhysRevLett.88.237902.

[54] Mark Whitney, Nemanja Isailovic, Yatish Patel, and John Kubiatowicz. "Automated generation of layout and control for quantum circuits". In: *Proceedings of the 4th international conference on computing frontiers (CF)*. Ischia, Italy: ACM, 2007, pp. 83–94. ISBN: 978-1-59593-683-7. DOI: 10.1145/1242531.1242546.

[55] Robert Wille, Oliver Keszocze, Marcel Walter, Patrick Rohrs, Anupam Chattopadhyay, and Rolf Drechsler. "Look-ahead schemes for nearest neighbor optimization of 1D and 2D quantum circuits". In: *21st asia and south pacific design automation conference (ASP-DAC)*. IEEE, Jan. 2016, pp. 292–297. DOI: 10.1109/aspdac.2016.7428026.

[56] Robert Wille, Aaron Lye, and Rolf Drechsler. "Exact reordering of circuit lines for nearest neighbor quantum architectures". In: *IEEE transactions on computer-aided design of integrated circuits and systems* 33.12 (Dec. 2014), pp. 1818–1831. DOI: 10.1109/tcad.2014.2356463.

[57] Katsuhisa Yamanaka, Erik D. Demaine, Takehiro Ito, Jun Kawahara, Masashi Kiyomi, Yoshio Okamoto, Toshiki Saitoh, Akira Suzuki, Kei Uchizawa, and Takeaki Uno. "Swapping labeled tokens on graphs". In: *Fun with algorithms. Fun 2014*. Vol. 8496: *Fun with algorithms*. Lecture Notes in Computer Science. Springer International Publishing, 2014, pp. 364–375. ISBN: 978-3-319-07890-8. DOI: 10.1007/978-3-319-07890-8_31.

[58] Louxin Zhang. "Optimal bounds for matching routing on trees". In: *SIAM journal on discrete mathematics* 12.1 (Jan. 1999), pp. 64–77. DOI: 10.1137/s0895480197323159.

[59] Alwin Zulehner, Alexandru Paler, and Robert Wille. "An efficient methodology for mapping quantum circuits to the IBM QX architectures". In: *IEEE transactions on computer-aided design of integrated circuits and systems* (June 7, 2018). ISSN: 1937-4151. DOI: 10.1109/tcad.2018.2846658.

[60] Alwin Zulehner and Robert Wille. "Compiling SU(4) quantum circuits to IBM QX architectures". In: *24th asia and south pacific design automation conference (ASP-DAC)*. Tokyo, Japan: ACM Press, 2019, pp. 185–190. ISBN: 978-1-4503-6007-4. DOI: 10.1145/3287624.3287704.